\documentclass[manuscript]{aastex}

\newcommand{\flux}{erg cm$^{-2}$ s$^{-1}$ \AA$^{-1}$}
\newcommand{\swift}{{\it Swift}}

\shorttitle{Ultraviolet Spectroscopy of Supernovae:\\ The First Two Years of  \swift\ Observations}
\shortauthors{Bufano et~al.}

\begin{document}

\title{Ultraviolet Spectroscopy of Supernovae:\\ The First Two Years of  \swift\ Observations }

\author{F. Bufano}
\affil{Universita' degli Studi di Padova, Dipartimento di Astronomia, Padova, IT 35122\\
NASA/Goddard Space Flight Center, Astrophysics Science Division, Code 661, Greenbelt, MD 20771, USA\\INAF-Osservatorio Astronomico di Padova, Padova, IT 35122 }
\author{S. Immler}
\affil{NASA/Goddard Space Flight Center, Astrophysics Science Division, Code 662, Greenbelt, MD 20771, USA \\
Department of Astronomy, University of Maryland, College Park, MD 20742, USA} 
\author{M. Turatto}
\affil{INAF-Osservatorio Astronomico di Catania, Catania, IT 95123, Italy}
\author{W. Landsman}
\affil{NASA/Goddard Space Flight Center, Astrophysics Science Division, Code 661, Greenbelt, MD 20771, USA }
\author{P. Brown}
\affil{Pennsylvania State University, Department of Astronomy \& Astrophysics, University Park, PA 16802, USA}
\author{ S. Benetti}
\affil{INAF-Osservatorio Astronomico di Padova, Padova, IT 35122, Italy }
\author{ E. Cappellaro}
\affil{INAF-Osservatorio Astronomico di Padova, Padova, IT 35122, Italy }
\author{S. T. Holland}
\affil{NASA/Goddard Space Flight Center, Astrophysics Science Division, Code 660.1, Greenbelt, MD 20771, USA \\
Universities Space Research Association, Columbia, MD 21044, USA \\
CRESST, NASA/Goddard Space Flight Center, Greenbelt, MD 20771, USA}
\author{P. Mazzali}
\affil{INAF-Osservatorio Astronomico di Padova, Padova, IT 35122, Italy \\
Max-Planck Institut fuer Astrophysik, 85748 Garching, Germany}
\author{P. Milne}
\affil{Department of Astronomy and Steward Observatory, University of Arizona, Tucson, AZ 85721, USA}
\author{N. Panagia}
\affil{Space Telescope Science Institute,Baltimore, MD 21218, USA\\
INAF-Osservatorio Astrofisico di Catania, Catania, IT 95123, Italy\\ 
Supernova Ltd,  Virgin Gorda, OYV \#131, British Virgin Islands}
\author{E. Pian}
\affil{INAF-Osservatorio Astrofisico di Trieste, Trieste, IT 34131, Italy}

\author{P. Roming}
\affil{Pennsylvania State University, Department of Astronomy \& Astrophysics, University Park, PA 16802, USA}
\author{ L. Zampieri}
\affil{INAF-Osservatorio Astronomico di Padova, Padova, IT 35122, Italy }
\author{A.A. Breeveld}
\affil{Mullard Space Science Laboratory, Holmbury St. Mary, Dorking, Surrey, 
RH5 6NT, UK}
\author{N. Gehrels}
\affil{NASA/Goddard Space Flight Center, Astrophysics Science Division, Code 661, Greenbelt, MD 20771, USA }

\begin{abstract}
We present the entire sample of ultraviolet (UV) spectra of supernovae (SNe) obtained with the Ultraviolet/Optical Telescope (UVOT) on board the \swift\ satellite during the first 2 years of observations (2005/2006). 
A total of  29 UV-grism and 22 V-grism spectra of 9 supernovae (SNe) have been collected, of which 6 are
thermonuclear (type Ia) and 3 core collapse (type Ibc/II) SNe.
All the spectra have been obtained during the photospheric
phase. After a comparison of the spectra of our sample with those in the literature 
(SNe 1992A, 1990N and 1999em), we confirm some degree of diversity in the UV emission 
of Type Ia SNe and a greater homogeneity in the Type II Plateau SN sample. 
Signatures of interaction between the ejecta and the circumstellar 
environment have been found in the UV spectrum of SN~2006jc, the only 
SN Type Ib/c for which UVOT grism data are available. 
Currently, \swift\ UVOT is the best suited instrument for early SN studies 
in the UV due to its fast response and flexible scheduling capabilities. However, in 
order to increase the quality of the data and significantly improve our understanding of the UV properties of SNe and to fully maximize the scientific potential of UVOT grism observations,
a larger investment in observing time and longer exposures are needed.

\end{abstract}


\section{Introduction}
Important information on  the explosion physics, progenitors, and environments of
supernovae (SNe) can be obtained from the analysis of ultraviolet (UV)
observations. 
The UV emission of SNe is strong in the early stage after explosion, 
when the  ejecta are hot and dense and the photosphere is located in the outer layers.
This makes UV data uniquely suited to study the outer layers of the
progenitor, and thus understand its structure before explosion. 
A better understanding of the UV properties of nearby SNe is also fundamental
for the cosmological use of high-$z$ SNe, both those presently discovered by ground-based 
wide-field surveys or by the {\sl Hubble Space Telescope} (HST), and those expected
to emerge from the next generation of space missions such as JWST and JDEM. 
This is because the optical and IR observations of high-$z$ SNe actually 
sample the rest-frame UV emission.
An adequate understanding of the UV emission of local SNe Ia and 
its time evolution will help to settle, for example, possible evolutionary biases with 
cosmic age in SNe Type Ia, allowing us to continue to use them as reliable 
distance indicators.  Recently \citet{FoleyHZ} and  \citet{Ellis08} indicated the paucity 
of high quality UV data of local SNe as the main limitation of their comparative 
analysis with the high-z SNe, concluding that an intense campaign of acquisition
of new UV data is urgently needed. 

Thirty years after early systematic SN observations with the {\sl
International Ultraviolet Explorer} (IUE), the sample of SNe observed in the UV
is still small. Currently, good quality UV data are available only for a
few events per SN type.
IUE took UV spectra of 23 SNe over its entire mission lifetime \citep{ulda}, while HST collected data for 8 and 11 SNe with FOS and STIS,
respectively, and low resolution spectra of 4 SNe with the ACS UV prism after
the failure of STIS. A complete review of UV observations of SNe was given by
\citet{Panagia2003,Panagia2007} pointing out the necessity of a larger UV sample of local SNe,
 in order to address the issue of similarity or diversity among both thermonuclear and core 
collapse SNe in the UV range.

A comprehensive study of the UV properties of SNe is the main goal of
an observing program established with the \swift\ satellite.  Because of the fast response and flexibility in terms of scheduling, a design requirement for the study of Gamma-Ray Bursts \citep{Gehrels}, \swift\ is uniquely suited for such a program. \swift\ also offers simultaneous coverage over a wide spectral range, spanning from the X-ray to the optical bands and thus providing information that cannot be obtained from ground-based telescopes or any other instrument alone.

In this paper we present the catalog of UV spectroscopic
observations of SNe obtained with \swift\ until December 2006. In \S\ 2, 
we present a brief overview of our present understanding of the origin of UV emission for
different SN types. A brief description of the capabilities 
of \swift, focusing on the UV wavelength range, and of the data reduction procedures
are given in \S\ 3. In \S\ 4, the observed SN sample
is presented, and the data analysis is described in \S\ 5.
Conclusions are drawn in \S\ 6. 

\section{Origin Of The Ultraviolet Flux}
In this section we summarize the scenarios
 proposed for the UV flux production for the two main SN classes (thermonuclear vs core collapse).
The available data gathered by IUE and HST seem to point to 
similar behaviors in the UV for Type Ia and Type Ib/c SNe \citep{Panagia2003}
which are therefore discussed together.

\subsection{Type I Supernovae}
The UV flux production of Type Ia and Ib/c SNe is only a small fraction of the total emitted flux. 
The UV photons, produced in the deep layers of the ejecta, are absorbed almost
entirely by a forest of transition lines of heavy elements such 
as Ti, Cr, Fe, Co and Ni in their single and doubly ionized species \citep{Pauldrach96},
and re-emitted at lower energies, emerging in the red part of the
spectrum where the opacity is smaller. 
In addition, because of the expansion of the SN atmosphere ($v=2,000$--$20,000~{\rm km\,s}^{-1}$), 
each line occupies a much more extended wavelength range than in a normal stellar atmosphere.
Therefore each UV photon is scattered from line to line by the gas, being progressively
redshifted until it finds a ''window'' to escape. This wavelength window is typically in the
red where the scattering lines are less numerous and optically thick  \citep{Fransson1994}.

  In particular, by modelling Type Ia SNe observed spectra, it has been 
found that the UV emission originates predominantly
from reverse fluorescence in iron group ions which converts photons from red to
blue wavelengths in the outermost layers of the SN ejecta. Here the UV opacity
is sufficiently small that photons can escape (\citealt{Lucy1999};
\citealt{Mazzali2000}). Therefore the emerging UV spectrum depends mainly on the
chemical composition of the outer layers of the ejecta. 
\citet{Lentz2000} suggested a dependence between the emitted flux and the 
progenitor metallicity. In their modeling they found that the flux was lower and the 
spectral lines blue-shifted for increasing metal content.
  On the other hand, in a recent work and similarly to \citet{Hoeflich} 
(who focused the analysis mainly on the
metallicity effects on the light curve shapes), \citet{sauer08} 
found that under certain conditions the UV flux may actually increase 
with metallicity. This behavior appears to reflect an enhanced probability of the
reverse-fluorescence process and a change in the ionization  fraction (``backwarming'' 
effect) as a consequence of higher metal content in the outer layers.  The distribution of
metals and their degree of mixing \citep{Blinnikov2000}, as well as the
kinematics of the layers in which the UV spectrum is produced, depend on
the explosion mechanism \citep{hille00}. Thus, the very early UV emission can provide unique
information about the progenitor nature and the explosion mechanism of SNe Ia.

\subsection{Type II Supernovae}
In Type II SNe, the UV emission represents the main fraction of the flux
emitted immediately after the explosion. In general during the collapse of the progenitor
core, a shock wave is
generated that propagates through the star and ejects the envelope.
As the shock wave emerges at the surface of the star,  very bright UV and X-ray
emission  flashes are expected (\citealt{Ensman}, \citealt{Blinnikov1993},
\citealt{Blinnikov1998}), lasting a brief period of time, from about 5 minutes
for a blue supergiant (BSG) progenitor to about one hour for the largest red supergiants
(RSG; \citealt{Levesque}). 
The short initial burst is then expected to be followed by a bright, post-breakout UV plateau phase 
lasting for about two days as a result of the competition  of two main effects: 
the cooling of the SN ejecta, due to the expansion and the emission of radiation
 which produces the shift of the spectral energy distribution (SED) to
longer wavelengths, and the strong decrease of the bolometric luminosity.

X-ray and UV emission from shock breakouts have been detected with \swift\ for the Type Ic SN~2006aj ( 
initially indicated as a X-Ray Flash, later recognized as a SN, \citealt{Campana},
\citealt{Waxman}) and the Type Ib SN~2008D (\citealt{Soderberg}, \citealt{Mazzali08D}, \citealt{Modjaz08D}).
In both cases the time duration and radii of the emitting sources are consistent with a shock 
breakout mediated by a dense stellar wind. 
Alternatively, a jet breakout scenario for SN~2008D \citep{Mazzali08D} and non-thermal 
synchrotron emission for SN~2006aj \citep{Ghisellini06aj} have been suggested.
  The emission from the shock breakout  was recently observed with GALEX
for a Type II-P SN (SNLS$\_$04D2dc, \citealt{Schawinski}), revealing as well that the UV flux received during the initial 
hours is associated to the radiative precursor of the shock, thus produced long before the latter reaches the surface.
Serendipitously detected UV plateau phases associated with the early breakout phase in two Type II 
Plateau SNe with the {\sl Galaxy Evolution Explorer} (GALEX) have also been reported \citep{Gezari08}.

A few days after the explosion, when the UV emission is already fading,  the SN becomes brighter 
in the optical domain and so more easily discoverable by robotic and other ground-based optical telescopes. 
With the lower temperature, as in Type I SNe, line blanketing becomes important because of the high
density of lines of low ionization stages of iron group elements (especially FeII  and FeIII; \citealt{Mazzali2000}; \citealt{Dessart2005}, 2006).
The UV is therefore directly linked to the metal content of the SN ejecta and to their
rapidly changing conditions at early phases. 

In the cases of SNe 1979C and 1980K, early spectra were characterized by  
an UV excess below 1500\,\AA\ and by emission lines of highly ionized species 
(e.g.\ N\,V, N\,III and Si\,IV; \citealt{Benvenuti82}; \citealt{Panagia1980}).
\citet{Fransson1984} suggested that such UV excess may originate  
from the interaction between the supernova ejecta and the pre-existing circumstellar 
material (CSM). Just behind the outward moving shock front energetic, thermal electrons 
(T$\sim10^9$\,K) are produced which give rise to soft X-ray emission. 
Inverse Compton scattering by these electrons becomes the main 
source of the energy of narrow-line UV emission, causing the ionization and excitation 
of the species in the outer layers of the SN envelope.
At the same time, the radiation from the shock wave dominates the continuum below 1500 \AA. 
The ejecta-CSM interaction has been used also to explain the UV emission
detected at late phases in some CC SNe (\citealt{Panagia1980}, \citealt{Fesen79c}, 
\citealt{Immler79C}, \citealt{Fransson2002}, 2005). UV emission can therefore
reveal the structure of the SN environment, which is directly linked to the final
phases of the pre-SN evolutionary history. 

\section{UV Spectroscopic Capabilities of \swift\ and Data Reduction }
\subsection{Overview}

\swift\ has three instruments that operate simultaneously: the Burst Alert
Telescope (BAT; \citealt{barth04}), the X-ray Telescope (XRT;
\citealt{burrows05}) and the Ultraviolet/Optical Telescope (UVOT;
\citealt{roming05}). The latter is the instrument of interest in this study.

\citet{PooleUVOT} have recently compared UVOT with the three other orbiting missions that 
have UV capabilities, i.e.\ the {\sl XMM-Newton} Optical Monitor 
(OM; \citealt{MasonXXM}), the  GALEX  \citep{MilliardGALEX,BianchiGALEX} and HST with the Wide-Field Planetary Camera 2 (WFPC2; 
\citealt{BurrowsHST}).  UVOT has a considerably higher spatial resolution ($<2''$ FWHM) than GALEX
and larger field of view ($17' \times 17'$) than HST. Paired with its fast response time ($\sim$hours), 
UVOT is a prime instrument to observe transient phenomena such as SNe. 
\swift\ also has flexible scheduling capabilities that allow frequent visits to a target and extensive
monitoring campaigns to follow a target's temporal evolution in unprecedented detail.  

UVOT is a Ritchey-Chr\'etien reflector telescope with a 30~cm primary mirror.
It uses a microchannel-intensified CCD (MIC) detector, a photon-counting device capable 
of detecting very faint signals. Through a filter wheel it provides imaging in six 
different bands (uvw2, uvm2 and uvw1 in the UV; u, b and v in the optical) spanning the
wavelength range 1700--6500\,\AA. Low resolution spectroscopy can be obtained with two grisms. 
A UV-grism provides spectra of moderate signal-to-noise ratio  per pixel  (S/N $\sim$10--15) 
for objects in the magnitude range $\approx 11$--15~mag over 1700--2900\,\AA\ with a resolution R$\sim$150,
comparable to low resolution IUE spectra. UV-grism spectra actually extend long-ward of
2900\,\AA, but above this wavelength the second-order overlaps the first order spectrum. 
The V-grism covers the wavelength 
range from 2800 to 5200\,\AA\ with a lower resolution (R$\sim$75), and thus
a slightly higher S/N ($\sim$15--20 per pixel),  down to fainter magnitudes 
(13--17~mag)\footnote{\small{See UVOT Grism Notes at
$http://heasarc.gsfc.nasa.gov/docs/swift/about\_swift/ uvot\_desc.html$ for a
comparison of the sensitive areas of the two different grisms.}}. The
contamination of the first-order UV spectra by zeroth-order from nearby 
sources can be minimized by using {\it clocked mode} observations. With this
mode, zeroth-order spectra only appear in the area covered by the grism, whereas
first-order spectra are dispersed off the edge of the grism. 

\subsection{Data Reduction\label{datareduction}} 

Because of the low-Earth orbit of \swift, most grism observations are composed of a sequence of multiple
exposures with  times $<10$~ks.
We have chosen to reduce and combine only the deepest of these observations that have a better S/N. 
First, each raw grism image is corrected for the {\it modulo-8} 
fixed-pattern noise, produced by the on-board algorithm used to read out the CCD at high rate and to
calculate the centroid position of the incoming photon splash \citep{roming05}.
From the zeroth-order position of the chosen source, called {\it anchor point},  with an automated tool the user
 selects the area in the grism image that contains the first-order spectrum, 
which is then extracted and calibrated in wavelength and flux. 
We use \swift\ calibration files  computed by using spectrophotometric standard stars,
 as discussed by \citet{Breeveld}  and made available by the UVOT instrument team on the \swift\ web site.
The above method provides reliable spectral extractions and calibrations for most UVOT
targets but not for SNe that are contaminated by strong emission from the host galaxies.
We therefore performed an interactive extraction for all SN
spectra using the APALL package in IRAF \footnote{\small{IRAF, the Image
Reduction and Analysis Facility, is distributed by the National Optical
Astronomy Observatory, which is operated by the Association of Universities for
Research in Astronomy, Inc. (AURA) under cooperative agreement with the National
Science Foundation (NSF).}}. 
While this interactive procedure is time consuming, it allows a considerable 
improvement compared to the automatic extraction routine.
An example is the grism spectrum of SN~2005cs, taken on 2005-07-06.29 UT, for which we were able to
recover the spectral range 2200--2600\,\AA\ (Fig.\ \ref{2005cs}) that was lost by Brown et al.\ 
(2007, their Fig.\ 3) because of contamination by the zeroth-order spectrum of a nearby field star.

Because the non-negligible dispersion of the zeroth-order spectrum gives a
poor anchor point reference, wavelength calibration is not a trivial task in grism spectroscopy.
In the case of  \swift\ there is the additional complication of a drift of up to 2$''$ between consecutive
exposures which prevents the use of existing, previous direct imaging to fix the
position of the source, and hence the wavelength anchor point. 
Therefore, such reference point is best determined by the centroid of the zeroth-order. 
The issue becomes even more troublesome for SNe, since their SED is different from 
that of the stars, typically used to calibrate the wavelength scale 
(usually White Dwarfs with black-body like spectra), and evolve with time. 
Consequently, the centroid of the zeroth-order spectrum can shift with respectof the UVOT spectra
to the reference value introducing an error of several Angstrom in the wavelength calibration. 
This problem affects the SN spectra presented in this paper as well as all previously published SN
spectra obtained with \swift.  

 The check of the wavelength calibration has been done by performing
an automatic cross-correlation of the UVOT spectra with ground-based spectra, where available. 
Only quasi-simultaneous spectra are considered ($|\Delta
t|\lesssim 1$ day)  excluding time evolution effects
in the calibration process. We measured anchor point offsets in the range from 0 to 66\,\AA\,
($\langle|\Delta\lambda|\rangle=16.7 \pm 17.8$\,\AA) finding that they are not systematic and 
do not depend on the SN evolutionary phase.
In \S\ \ref{discussion}, we report the estimated offset values for each SN spectrum.  

  The wavelength scale of the UVOT grisms varies nonlinearly with detector 
position, and this effect can be significant  for large ($>$100 pixel) 
shifts.   In order to overcome this problem, all \swift\ spectra were taken
with the SN in the same position on the detector ($<$50 pixels), by mean of a  
second slew.  
Using a model of the nonlinearities with the detector position, 
we estimated a residual of 1.5 A between 2500 A and 5000 A.  \\

Finally in order to obtain the absolute flux calibration of each spectrum, 
we compared the simultaneous \swift\ broad-band magnitudes to the synthetic ones
obtained convolving each spectrum with the transmission curve of the considered filter. 
Subsequently,  we scaled the spectrum for the average value found for all the
available filters. Broad-band magnitudes are reported in the following spectrograms as 
filled dots at the corresponding Vega-effective wavelengths \citep{PooleUVOT}.  
The UVOT photometry collected
in these first two years was taken from \citet{BrownUV}, Milne et al.\ (in preparation) for SNe Ia, 
\citet{DessartUV} for SN~2005cs and 2006bp [updated by \citet{Brown05cs} and \citet{Immler06bp},
respectively], and Modjaz et al.\ (in preparation) for SN~2006jc.  Recently \swift\ photometry 
of SN~2005cf has been presented in the comprehensive work by \citet{Wang05cf}.    
 
 \section{Supernova Sample}
Target selection criteria were used to maximize the scientific return of the
UVOT observations. A SN suitable to be observed by UVOT (either in imaging or grism mode) should have the following properties: (i) young, i.e., discovered before maximum in the optical, (ii) nearby (z $\lesssim $ 0.01), (iii) low interstellar extinction along the line of sight ($A_V \lesssim 0.5$~mag), (iv) favorably located (distance $>8''$ from the host galaxy nucleus or bright field stars), and (v) low impact on \swift\ gamma-ray burst (GRB) studies (Sun angle $>90\degr$; angle from recent GRB $>60\degr$). These criteria led to the observation of 32 SNe with \swift\ UVOT during the first two years (2005--2006). For grism observations, an additional constraint on the target brightness (optical $< 15$~mag) was adopted that limits the sample of spectroscopic observations to 9 SNe. A brief introduction to these 9 individual SNe is given below; a summary of their main properties and the journal of UVOT grism observations are reported in Tables \ref{SNsample} and \ref{observation}, respectively.

{\bf SN~2005am} was discovered by \citet{Martin05am} in images taken on 2005-02-22.73 and 
2005-02-24.58~UT and confirmed one week later by \citet{Yamaoka05am}.  
An optical spectrum obtained on 2005-03-03.29~UT showed it to be a SN Ia, one or two
weeks before maximum \citep{Modjaz05am}.  \swift\ grism observations were taken
between 2005-03-08 and 2005-03-24~UT. Five spectra were obtained both with the UV and the
V-grism, for a total of 10 spectra. A preliminary analysis was presented by
\citet{brown05am}. Of all UV-grism exposures, only the spectrum taken on JD
2453439.1 turned out to be useful because of contamination with a nearby
field star. The light curve shows that the $B$-band maximum
occurred on JD 2453438 $\pm$ 1 day \citep{brown05am}.

{\bf SN~2005cf} was discovered with the Katzman Automatic Imaging Telescope (KAIT) in unfiltered image taken on 2005-05-28.36~UT \citep{Puckett05cf} and classified as a SN Ia about ten days before maximum light on 2005-05-31.22~UT \citep{Modjaz05cf}. Twelve UV
grism and ten V-grism observations were obtained with \swift\, between 2005-06-04 and 
2005-06-29~UT.  The observations are the most detailed campaign of a SN obtained
with UVOT so far. A $B$-band maximum on JD 2453534.0 $\pm$ 0.3 day \citep{pastorello05cf} is adopted in this paper.

{\bf SN~2005cs} was discovered on 2005-06-28.90~UT by \citet{Kloher} and
classified as a young SN~II by \citet{Modjaz05cs}.  On the basis of
pre-discovery limits, \citet{pastorello05cs} estimated that the explosion
occurred on 2005-06-27.5~UT (JD=2453549 $\pm$ 1). \swift\ observations began on 2005-07-03
and ended on 2005-07-19~UT. A total of 6 UV-grism and 2 V-grism spectra were
collected which have been presented and discussed in \citet{Brown05cs}.

{\bf SN~2005df} was discovered by \citet{Evans05df} on 2005-08-4.62~UT.  
\citet{salvo} classified it as a peculiar SN~Ia a few days before maximum light
based on a spectrum  obtained on 2005-08-05.83~UT with the Australian National University (ANU) 2.3-m telescope (wavelength range 390--700 nm). 
No light curve has yet been published to date and no UVOT photometry was collected 
in the optical bands, leaving the epoch of $B$-band maximum unconstrained.
Unfiltered observations reported on the
SN Web\footnote{See http://astrosurf.com/snweb2/2005/05df/05dfMeas.htmhttp://astrosurf.com/snweb2/2005/05df/05dfMeas.htm} suggest
that SN~2005df peaked around 2005-08-18.1~UT (JD 2453600.6).
UV-grism observations were obtained on four epochs, starting on 2005-08-11~UT.  

{\bf SN~2005hk} was discovered by the Lick Observatory
Supernova Search (LOSS) on 2005-10-30.25~UT \citep{burket05hk}.  
Initially, SN~2005hk was classified as a SN~Ia one or two weeks before maximum
light, similar to SN~1991T \citep{serduke}. The unusually low expansion velocity
(6000--7000 km\,s$^{-1}$) suggested that SN~2005hk was a SN~2002cx-like
object which was confirmed by a spectrum taken by the Carnegie Supernova Project
on 2005-11-23.2~UT (\citealt{Chornock05hk}, \citealt{Vallery05hk}, \citealt{Sahu05hk},
\citealt{phillips05hk}). Only one UV-grism spectrum was taken with \swift\ on 2005-11-08~UT,
 about 3 days before maximum in the $B$-band (JD 2453685.1 $\pm$ 0.5 days,
\citealt{phillips05hk}) at which time it was already rapidly fading in the UV \citep{BrownUV}. 

{\bf SN~2005ke} was discovered on 2005-11-13.33~UT by KAIT \citep{Puckett05ke}
and later classified as an under-luminous SN~Ia from the presence of the
characteristic 4200\,\AA\  Ti II band \citep{patat05ke}.  We assume JD 2453699
$\pm$ 2 days as the epoch of the $B$-band maximum on the basis of \swift/UVOT
photometric observations \citep{Immler05ke}. One V-grism and three
UV-grism spectra of low S/N before maximum were collected.  
Therefore, we have no insight into the spectroscopic UV properties or nature of the UV excess seen in the light curves starting 15 days after maximum \citep{Immler05ke}.   

{\bf SN~2006X} was discovered independently by Suzuki and Migliardi on 2006-02-07.10~UT
 \citep{circ06x}. Soon thereafter, \citet{Quimby06x} classified it as a
very young SN~Ia. The spectral features are similar to those of SN~2002bo
\citep{Benetti02bo} 1--2 weeks before maximum light, but with a red
continuum.  Prompt imaging obtained with \swift\ \citep{Immler06x} revealed only
weak emission at short wavelengths, likely due to strong interstellar
absorption in the host galaxy. As a consequence, only V-grism spectroscopy was 
obtained about 10 and 1 day before maximum (occurred on 2006-02-19.93~UT, JD 2453786.2,
 \citealt{Wang06x}). Variability in narrow features of the Na~I D lines has revealed 
the existence of CSM suggesting that the white dwarf was accreting material from a 
companion star during the red-giant phase \citep{Patat06x}.

{\bf SN~2006bp} was discovered on 2006-04-09.60~UT \citep{nakano06bp}. 
\citet{ImmlerAtel06bp} classified it as a young SN~II on the basis of the $B-V$
and $U-B$ colors  measured with UVOT \citep{BrownUV}. This classification was
confirmed by a HET spectrum taken two days later \citep{QuimbyCBET06bp}. The
spectrum of SN~2006bp showed a blue continuum with a narrow emission line
consistent with rest-frame H${\alpha}$. Seven UV-grism spectra were taken
between 3 and 14 days after the explosion, which is assumed to have occurred around 2006-04-09~UT (\citealt{QuimbyCBET06bp}, \citealt{Immler06bp}). Strong contamination due to
bright field stars affected the early spectra, but using the 'clocked' grism mode (see \S\ \ref{datareduction}), two useful follow-up UV spectra were obtained. A single V-grism observation was also taken on 2006-04-21~UT.

{\bf SN~2006jc} was discovered in UGC 4904 by \citet{nakano06jc} on 2006-10-9.75~UT.
 An upper limit was obtained earlier on 2006-09-22~UT, but a possible precursor
eruption was registered two years prior to the SN (\citealt{nakano06jc},
\citealt{pastorello06jc}). Because of the lack of H features and the presence of
narrow He I emission lines superimposed on a broad-line spectrum, SN~2006jc was
defined as a peculiar Ib object (e.g., \citealt{crotts06jc},
\citealt{fesen06jc}, \citealt{benetti06jc}, \citealt{Modjaz06jc},
\citealt{pastorello06jc}, \citealt{foley06jc}). \citet{benetti06jc} highlighted
the similarity with other rare events, namely SN~1999cq \citep{Matheson99cq} and
2002ao \citep{Filippenko02ao}. According to \citet{Immler06jc}, the SN
explosion occurred on 2006-09-25~UT ($\pm$ 5 days).  Recently, \citet{Pastorellofamily}
discussed the physical properties of this class of objects. For SN~2006jc, we
obtained three UV-grism and three V-grism spectra, but only one of them was not
contaminated by the zeroth-order spectrum of a field star.

\section{Results and Discussion}\label{discussion}
In this section we group the targets into two samples: 6 thermonuclear and 3 core
collapse SNe for a total of 9 SNe for which   \swift/UVOT spectra are available.
For each SN we study the spectral evolution before and after maximum light while focusing on the
main features and highlighting similarities and/or differences.  As shown in
Tab.\ \ref{observation}, \swift\ collected a total of 41 UV-grism spectra and
24   V-grism spectra, of which  29  and 22, respectively, turned out to be
useful for our purposes.  
In col.~5 of Tab.\ \ref{observation}, we list  the total  exposure time of 
each spectrum, resulting from either a single long exposure or a series of co-added
shorter exposures (cfr.\ \S\ \ref{datareduction}) and corrected for dead-time. 
The exposure times relative to SN~2005am are slightly different from those published by  
\citet{brown05am}, who reported the total elapsed exposure time.

\subsection {Thermonuclear SNe}

\subsubsection{SN~2005cf}
\begin{figure}
\plotone{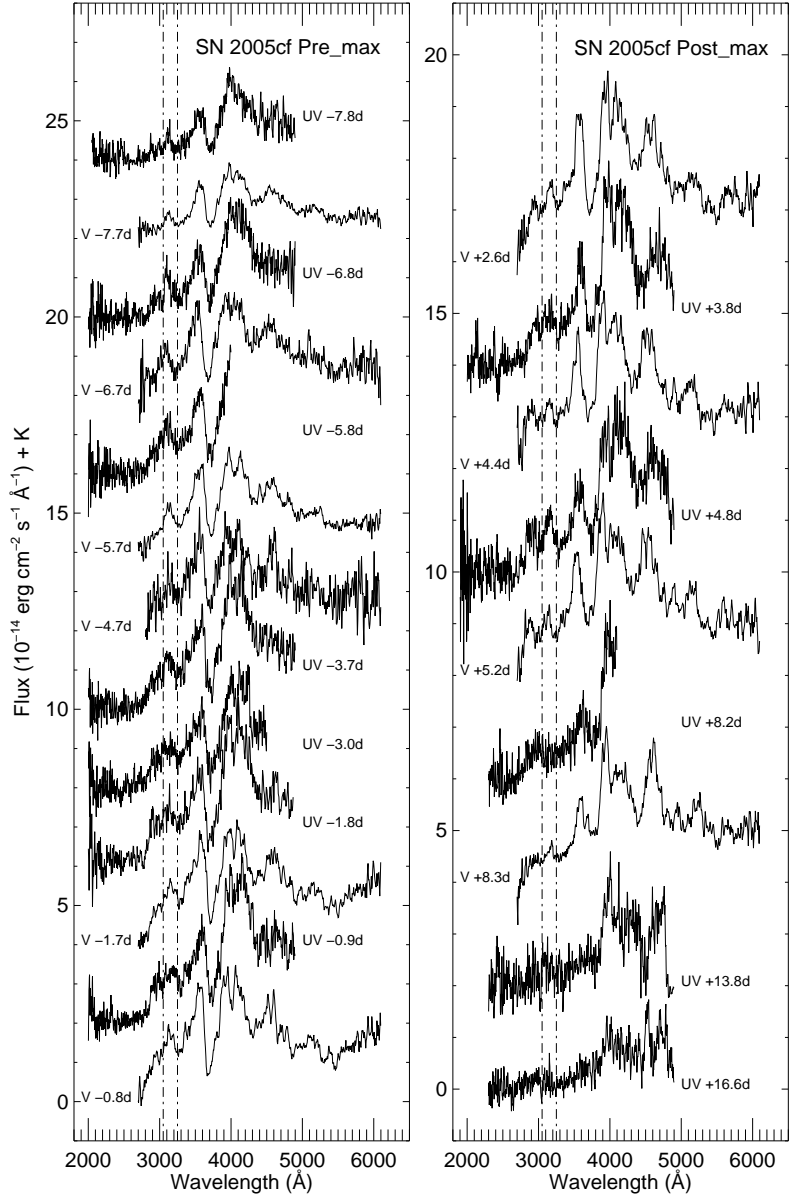}
\caption{Spectral evolution of SN~2005cf (Type Ia).  UV and V-grism spectra are plotted together in pre-maximum ({\it left panel}) and post-maximum ({\it right panel}) epochs. Phases are with respect the $B$ maximum light. The spectra are spectrophotometric calibrated by using the simultaneous \swift\ photometry and displaced by multiples of $2 \times 10^{-14}$ \flux. Vertical lines indicate the two main UV absorption features at $\sim3050$\,\AA\ and  $\sim3250$\,\AA\ (see text).  \label{05cfevol}  }
\end{figure}

Of all SNe observed by \swift, SN~2005cf has the most detailed time sequence of UV 
spectra. The UV and V-grism spectra are displayed in Fig.\ \ref{05cfevol}.
The UV spectra have a typical S/N $\sim$10  per pixel  and provide unique information about the SN
behavior in the range between 2000 and 3500\,\AA. 
The MgII $\lambda$2798 absorption doublet redshifted to about $2700$\,\AA\ is difficult 
to identify because its blue wing is suppressed by heavy metal line blanketing.
Because of the limited S/N, SN~Ia spectra are limited blue-ward of this wavelength.
A number of features are easily recognizable at longer wavelengths:
a strong CaII H\&K doublet at $\sim$3900 \AA\ (possibly contaminated by SiII
$\lambda$3858),  SiII $\lambda$4130 and $\lambda$4580, MgII $\lambda$4481, and 
blends of FeII and FeIII lines in the red part of the spectra.
Identified UV-optical lines are marked in Fig.\ \ref{05cf_evol}. 

\begin{figure}
\plotone{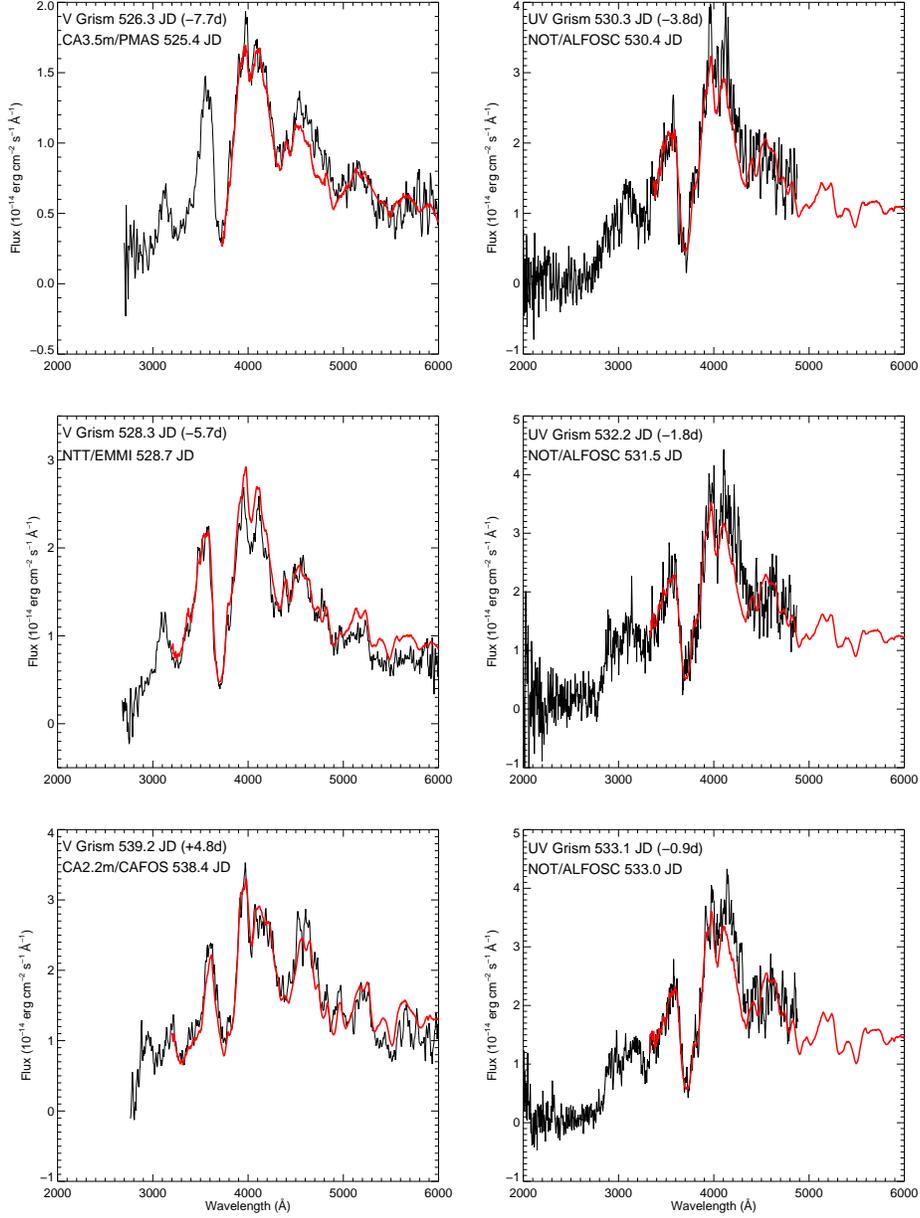}
\caption{Superposition of the UV and V-grism spectra of SN~2005cf (black line) with the corresponding
quasi--simultaneous, ground-based spectra (red line, from \citealt{Garavini05cf}), after wavelength offset correction (Tab.\ \ref{tabellashift}). Epochs (JD +2453000) and phases after the $B$-maximum light are indicated at the top left corners.  \label{multiple}}
\end{figure}

\begin{figure}
\plotone{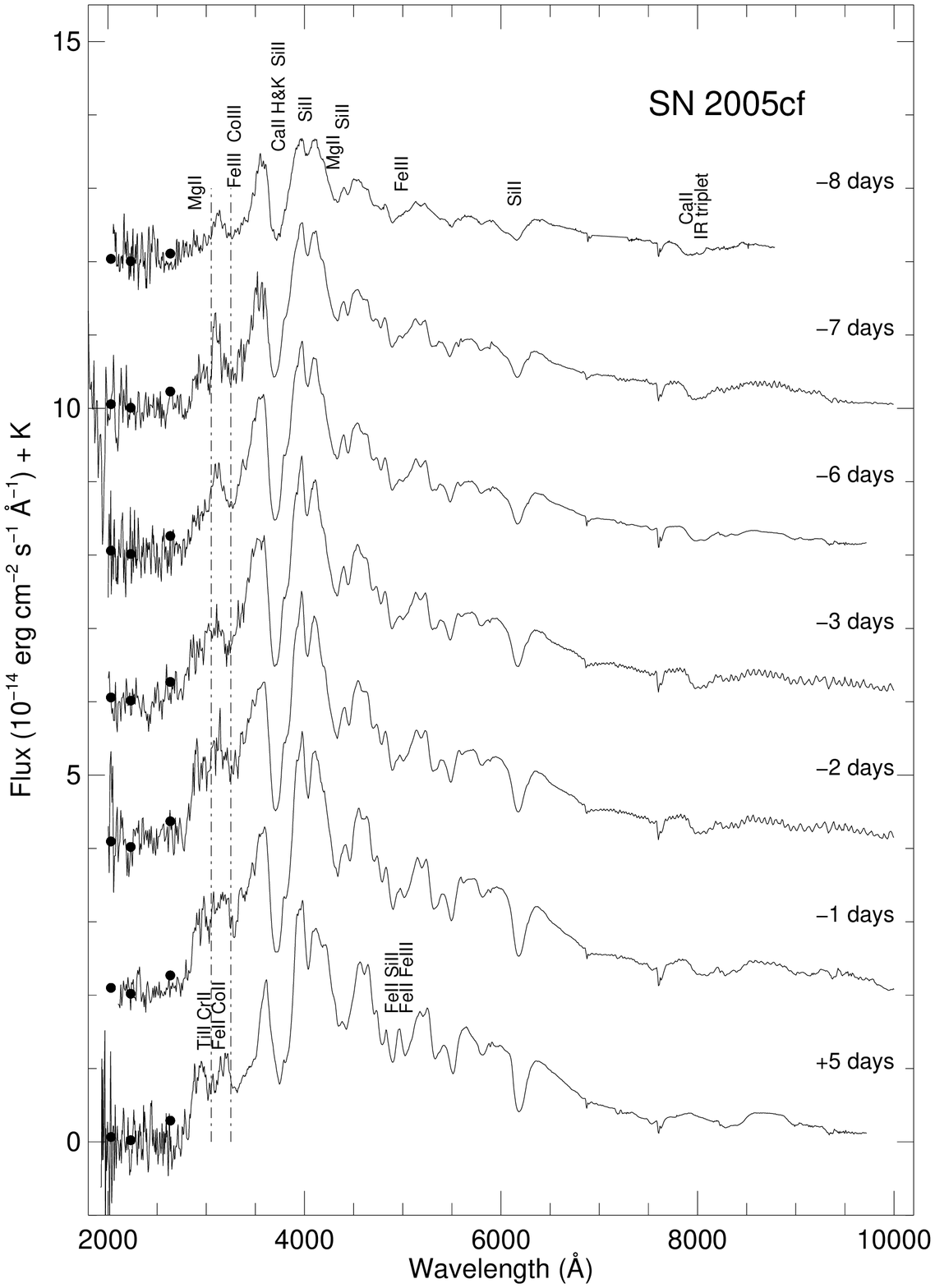}
\caption{Evolution of the UV-optical spectrum of SN~2005cf. The UV parts are from UVOT spectroscopy (smoothed with a boxcar of 5\,\AA), while optical spectra are from \citet{Garavini05cf}. Ordinates refer to the bottom spectrum while other spectra are vertically shifted by multiples of $2 \times 10^{-14}$ \flux. Identifications of the most conspicuous lines are also marked. UVOT broad-band photometry is indicated by black points at the effective wavelengths of the bands. \label{05cf_evol}}
\end{figure}

A careful analysis of the spectra at different epochs shows irregular shifts in wavelength. These are not intrinsic but caused by problems in determining the 
anchor point of the wavelength scale (see \S\ \ref{datareduction}). 
Where possible, the anchor point offsets have been determined by   a cross-correlation  with
quasi-simultaneous ground-based spectra \citep{Garavini05cf}, listed in 
Tab.\ \ref{tabellashift} along with the measured offsets. The offsets are typically on
the order of tens of \AA\ and do not show a systematic trend with SN epoch.
In Fig.\ \ref{multiple} we overplot part of the UVOT spectra with
the corresponding ground-based spectra. After wavelength correction, good
overall agreement is obtained for the SEDs and line intensities for both grisms.
UV-grism spectra are reliable long-ward of $\sim2900$\,\AA\ since 
second-order contamination is negligible because of the low flux below $\sim2800$\,\AA.
Around 4500\,\AA, V-grism spectra taken at $-7.7$d and +4.8d from
maximum show a (non-systematic) excess of about 10--15\% with respect to the corresponding
ground-based spectra.

Following \citet{FoleyUV}, we have tried to estimate the ''UV-ratio'' 
($\mathcal{R}_{UV}$) for the spectra near-maximum
($-3 < t < 3$~d), that were corrected for the anchor point offset (Tab.\ \ref{tabellashift}).
$\mathcal{R}_{UV}$, defined as the flux ratio $f_{\lambda} (2770\,{\rm\AA})/f_{\lambda} (2900\,{\rm \AA})$,
is used as an indicator of the SN UV spectral shape.
\citet{FoleyUV} found a correlation with the SN luminosity,
with bright (slowly declining) SNe Ia characterized by a small UV-ratio
(ranging between about 0.23 and 0.33).
For the  $-1.8$~d spectrum (2005-06-10.7~UT), the measurement is uncertain
because of the low signal from the short exposure time.
For the  $-0.9$~d spectrum (2005-06-11.6~UT), we find  
$ \mathcal{R}_{UV} = 0.23\pm0.14$. With a $\Delta {\rm m}_{15} ({\rm B})=  1.12\pm0.03~$mag,
SN~2005cf confirms the trend reported by \citet{FoleyUV}. 

The combined UV-optical spectra between $-8$d and +5d are shown in Fig.\ \ref{05cf_evol}.
An increase in flux at all wavelengths is observed for SN~2005cf as it approaches the maximum while the SED becomes redder.
The broad P-Cygni lines, among which the strongest are CaII and SiII,
become progressively narrower. \citet{Garavini05cf} made
a detailed analysis of the optical and near IR evolution of SN~2005cf, confirming
that it followed the evolution of a normal SN~Ia. The early spectra plotted in Fig.\ \ref{05cf_evol} 
show clear signatures of high velocity features (HVF, \citealt{MazzaliHVF}) in CaII
(both H\&K and the IR triplet) and SiII, with velocities of 24,000 km\,s$^{-1}$ and
19,500 km\,s$^{-1}$ respectively. These remain well detectable up to maximum light
\citep{Garavini05cf}.  

Focusing our attention on the UV range between 1800 and
3500\,\AA, which represents the novel contribution of \swift, we note two
features at $\sim$3050\,\AA\ and $\sim$3250\,\AA\ (cfr. Fig.\ \ref{05cfevol}). 
The 3250\,\AA\ feature is relatively broad before maximum and becomes narrower when
approaching it. Past maximum, the line profile becomes again broader in the red wing
likely caused by a blend with other absorptions.
The minimum near 3050\,\AA\ becomes  strong and well defined close to maximum. 
\citet{BranchUVspectra} suggested that the spectral features  seen in the 2750--3450\,\AA\ 
range of the early spectra are produced by blends of FeII and CoII lines. These ions 
are not responsible for the conspicuous features in the optical spectrum, 
which is shaped by neutral or singly ionized ions such as  OI, MgII, SiII, SII and CaII. 
 In particular, the 3250\,\AA\ minimum has been
explained with blue-shifted CoII absorption (rest wavelength 3350--3500\,\AA;
\citealt{Branch85model}) while the  $3050$\,\AA\ one with FeII absorptions 
\citep{BranchUVspectra}. The identification of Fe and Co is
consistent with the theoretical prediction that the light curve is powered by
the radioactive decay chain $^{56}$Ni--$^{56}$Co--$^{56}$Fe. 
Based on a study of the UV spectra of SN~1992A taken around maximum both with
IUE and  HST/FOS, \citet{Kirshner92A} confirmed that the spectrum in this range is 
shaped by blends of Fe-peak element lines.
In particular, they found a significant contribution  from CrII and FeII at $3050$\,\AA\
and from CrII, MnII and FeII at $3250$\,\AA, but excluded a contribution from 
CoII and NiII at the earliest epochs because the freshly synthesized
Ni and Co are confined to the inner region. Similar results were recently reported by
\citet{sauer08} who also highlighted the importance of the contribution of TiII in
forming the absorption at $3050$\,\AA\ and of the double ionized Co and Fe to the 3250 \AA\ feature, especially 
at epochs around the maximum.  
In general, strong blending makes line identification in the UV difficult. 

\subsubsection{SNe 2005df and 2005am}
\begin{figure}
\plotone{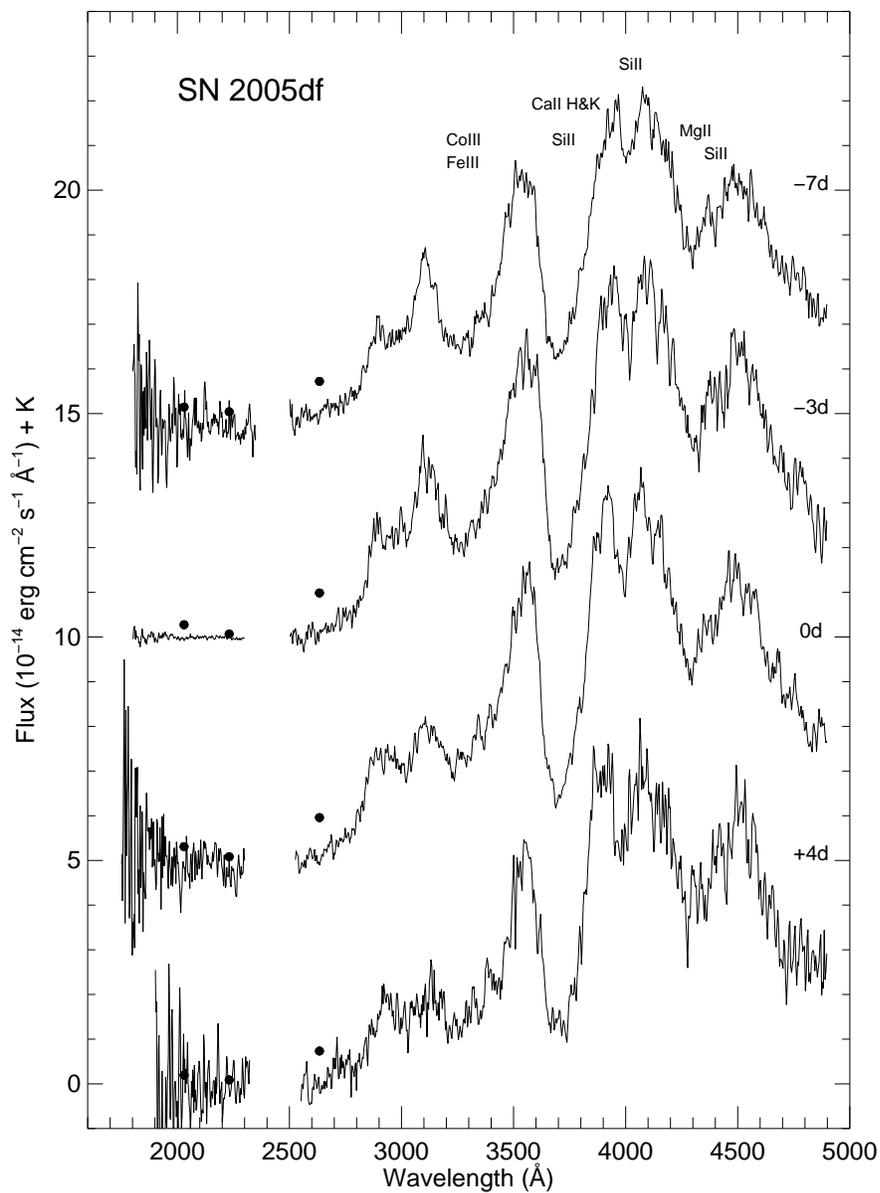}
\caption{ Spectroscopic evolution of SN~2005df (Type Ia).  
The gaps between 2300--2500\,\AA\  are due to contamination of the zeroth-order spectrum of a field star. Flux scale is as in the previous figure with spectra shifted by multiples of  5$\times10^{-14}$ \flux.  Wavelength calibration could not be checked against ground based spectra. UVOT broad--band photometry is indicated by black points at the effective wavelengths of the bands.
\label{05df}} 
\end{figure}

\begin{figure}
\plotone{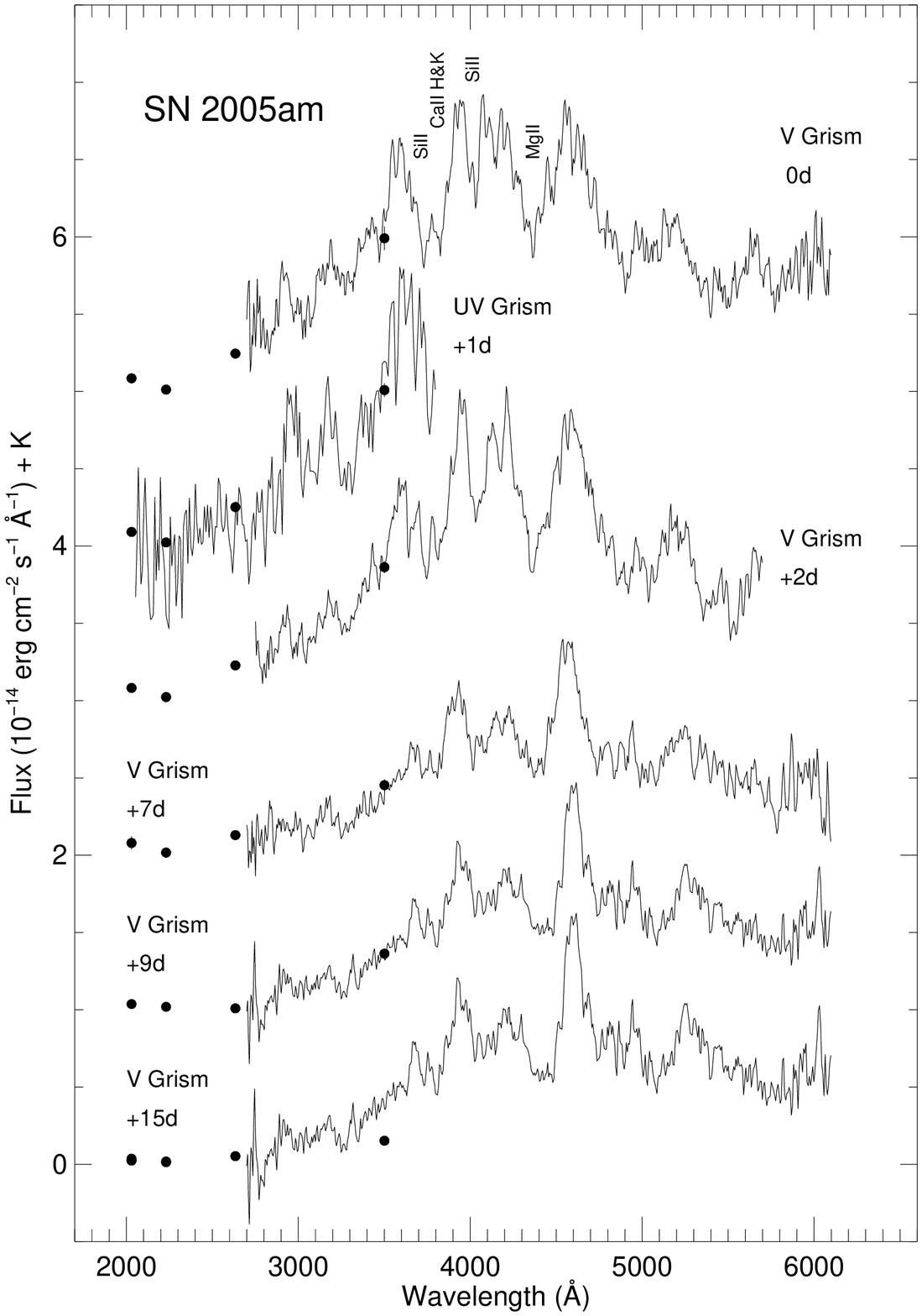}
\caption{UVOT grism spectra of SN~2005am (Type Ia). Each spectrum is labeled with its epoch in days after $B$ maximum light and, for clarity, shifted in flux by multiples of $1\times10^{-14}$ \flux. Spectra are scaled to match the \swift\ photometry and smoothed using a boxcar of 3 pixels ($\sim10$\,\AA). The wavelength scale could not be checked against quasi-simultaneous ground-based spectra.
UVOT broad--band photometry is reported as black points at the effective wavelengths of the bands. \label{05am} }
\end{figure}

The spectral evolution of SNe 2005df and 2005am is shown in Figures \ref{05df} 
and \ref{05am}, respectively.  The lack of available ground-based optical spectroscopy
makes the wavelength scale of the grism data uncertain  and possible $\mathcal{R}_{UV}$
measurements not reliable. 
Most UV-grism spectra for SN~2005df were collected before maximum. 
The gap between 2300--2500\,\AA\ 
in the spectrograms of Fig.\ \ref{05df} is due to contamination by a field star. 
Similarly to SN~2005cf, two broad minima at about 3050\,\AA\ and 3250\,\AA\ are present 
and show the same temporal evolution. 
Only post-maximum spectra are available for SN~2005am. The SN shows deep absorption
features at maximum which become weaker about a week later, such as the strong
SiII $\lambda$4130 absorption at about $4030$\,\AA. The SiII $\lambda$3859 and
CaII H\&K absorption lines, usually blended in a broad feature at
$\sim$3600--3800\,\AA\ (e.g. SN~2005cf), are narrow and appear resolved. The
persistence of the separation of the two features two weeks after maximum
seems to exclude the presence of HVF of CaII H\& K and supports the
identification of the blue component as the SiII line. Seven days after maximum, the 
optical spectrum is characterized  by the usual broad absorption lines, including SiII
($\lambda$4130, $\lambda$5051 and $\lambda$5972), MgII ($\lambda$4481) and
 numerous blends of FeII and FeIII lines. The minima near 3050\,\AA\ and 3250\,\AA\ qualitatively follow the evolution of SN~2005cf.  

\subsubsection{Comparison of SN~Ia around maximum light}
\begin{figure}
\plotone{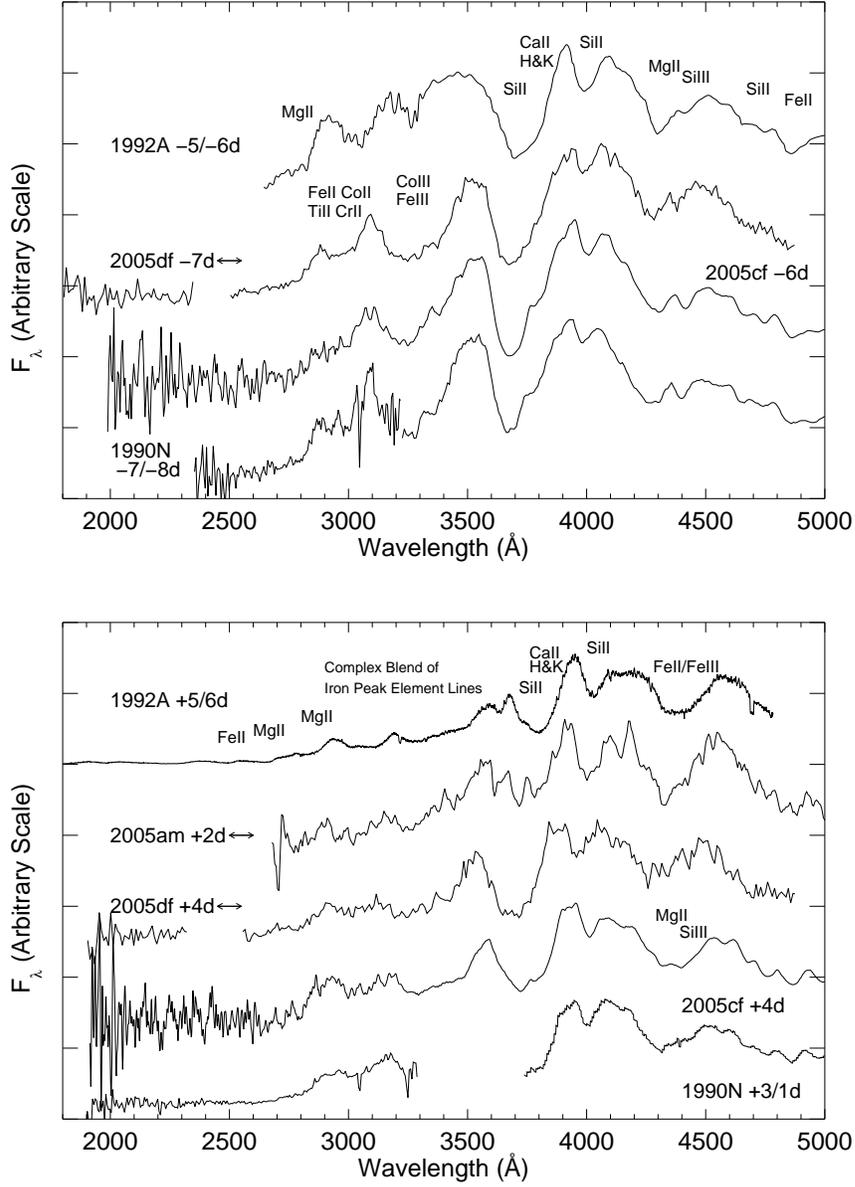}
\caption{{\footnotesize Comparison of UV-optical spectra of Type Ia SNe 1 week before
({\it upper panel}) and about 5 days after maximum  ({\it lower panel}).
The UV spectra of the nearby
SN~1992A were obtained with IUE for the former epoch (-5 days) and with HST/FOS
for the latter one ($+5$~d, \citealt{Kirshner92A}). They are combined with CTIO
optical observations at $-6$ and $+6$~d, respectively.  
UV spectra of SN~1990N were taken with IUE on day $-7$
and $+3$ \citep{ulda}, while optical counterparts were obtained
on day $-8$ with CTIO \citep{Leibundgut90n} and +1 with Asiago 1.82m telescope
\citep{Mazzali90N}. 
Horizontal double-arrows, close to the SN label, indicate possible shifts in the wavelength calibration.
The epoch of each spectrum relative to the maximum light is indicated in figure. The spectra are corrected for Galactic extinction \citep{Schlegel} and reported to the galaxy restframe, then scaled to the same magnitude and vertically shifted for clarity. }\label{conf_Ia} }  
\end{figure}

In Fig.\ \ref{conf_Ia} we compare the spectra of five SNe Ia, 1992A, 1990N,
2005df, 2005cf, and 2005am at two epochs, one week before (upper panel) and about
5 days after maximum (lower panel). 
The UV spectra of SN~1992A taken both with IUE and HST are of high S/N and allow 
a useful comparison of the performance of various instruments. 
The UV spectrum is produced in the outer layers because of the large opacity and can be 
strongly modified by small changes in the physical conditions (chemical composition, density
structure, etc.) which may have negligible impact at other wavelengths. 
Our comparison confirms that UV spectra
of SNe Ia can differ significantly, both before and  after maximum,
as shown also by \citet{sauer08} and \citet{FoleyHZ}. 
Before maximum, SNe 2005df and 2005cf appear more similar to SN~1990N than to SN~1992A. 
While red-ward of the CaII H\&K absorption the four SNe are quite
similar, in the UV  they are markedly different. In SN~1992A
the $3250$\,\AA\ absorption trough is narrow and lacks the extended red wing.
The relative ratio SiII $\lambda$4130 / SiIII $\lambda$4560
is much larger in SN~1992A than in other objects, which 
indicates lower temperature and ionization \citep{Mazzali90N}.
In general, all features of SN~1992A seem broader, as if washed out by larger 
expansion velocities: a clear example is the SiIII $\lambda$4560 line. 

\begin{figure}
\plotone{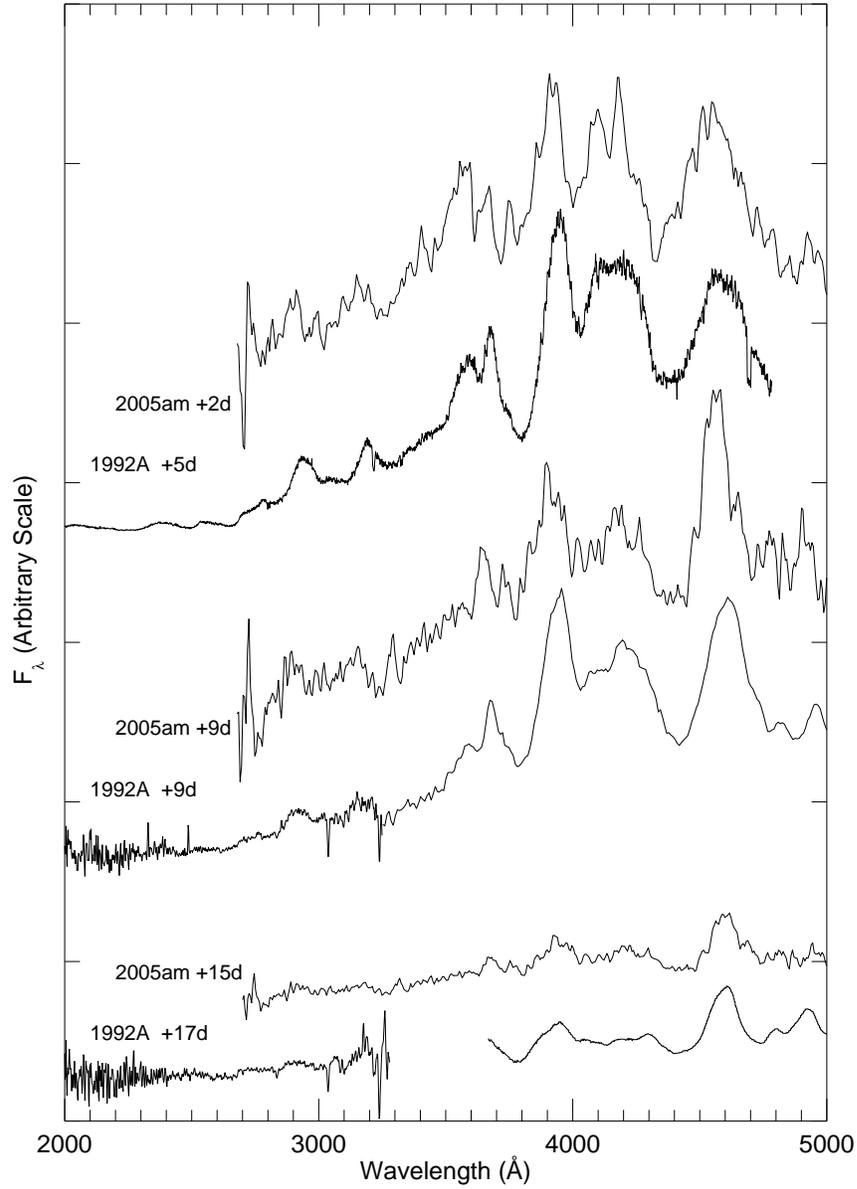}
\caption{Comparison of SN~1992A and 2005am spectra at different stages of their evolution. Spectra are corrected for Galactic extinction \citep{Schlegel} and reported to the host galaxy restframe. For clarity, fluxes are scaled to the same magnitude and vertically displaced.\label{conf05am92a} }  
\end{figure}

The differences between SN~1992A and all other SNe persist also 
after maximum. Again, the expansion velocities seem higher
so that the MgII and SiIII visible in 2005cf at about 4300 and 4400\,\AA\
\citep{Garavini05cf} are completely blended
to form a broad feature, which was attributed to FeII and FeIII lines by \citet{Kirshner92A}.
SN~1992A also shows an absorption at $\sim3650$\,\AA. This can be possibly recovered only in
the noisy spectrum of SN~2005am, which shares with SN~1992A the same
photometric behavior ($\Delta$m$_{15}(B)\backsimeq$ 1.4).
In Fig.\ \ref{conf05am92a}, the comparison between the two objects reveals an overall 
similarity with exception of an apparent slower expansion velocity for SN~2005am.

\subsubsection{ Peculiar SNe 2005ke and 2005hk}
\begin{figure}
\plotone{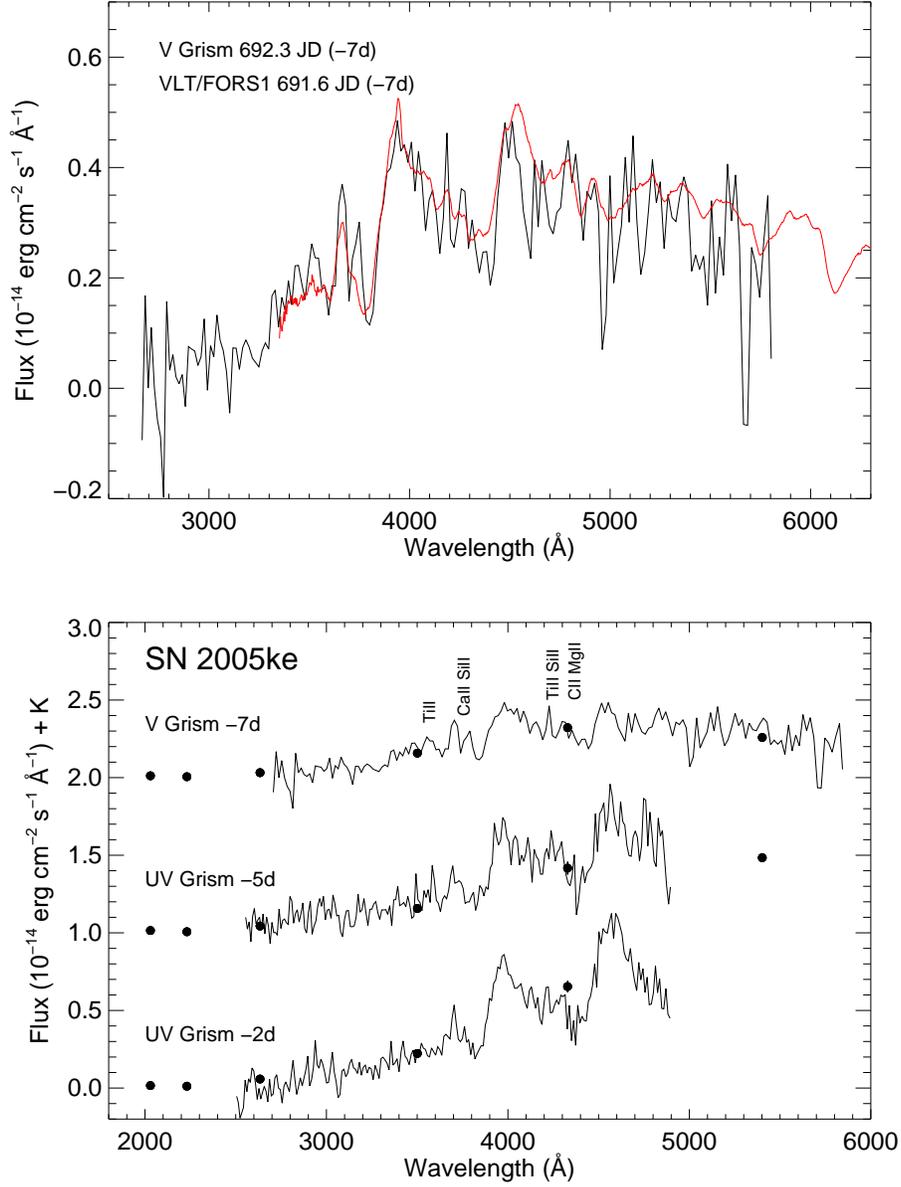}
\caption{SN~2005ke. {\it Upper panel:} Superposition of the UVOT V-grism spectrum taken on day $-7$ (black line) and a quasi-simultaneous ground-based telescope (VLT/FORS1, courtesy of F.Patat)  spectrum  (red line). A wavelength offset of $+41$\,\AA\ was measured. {\it Lower panel:} Spectral evolution of the Type Ia SN~2005ke. Spectrophotometric calibration was applied to the spectra by using simultaneous \swift\ photometry (black points). Spectra are shifted in flux by  multiples of $1 \times 10^{-14}$ \flux\ and smoothed ($\sim$15\,\AA). The wavelength calibration was performed using ground-based spectra for the V-grism spectrum and from the location of apparent TiII features (see text). \label{sn2005ke}}
\end{figure}

\begin{figure}
\plotone{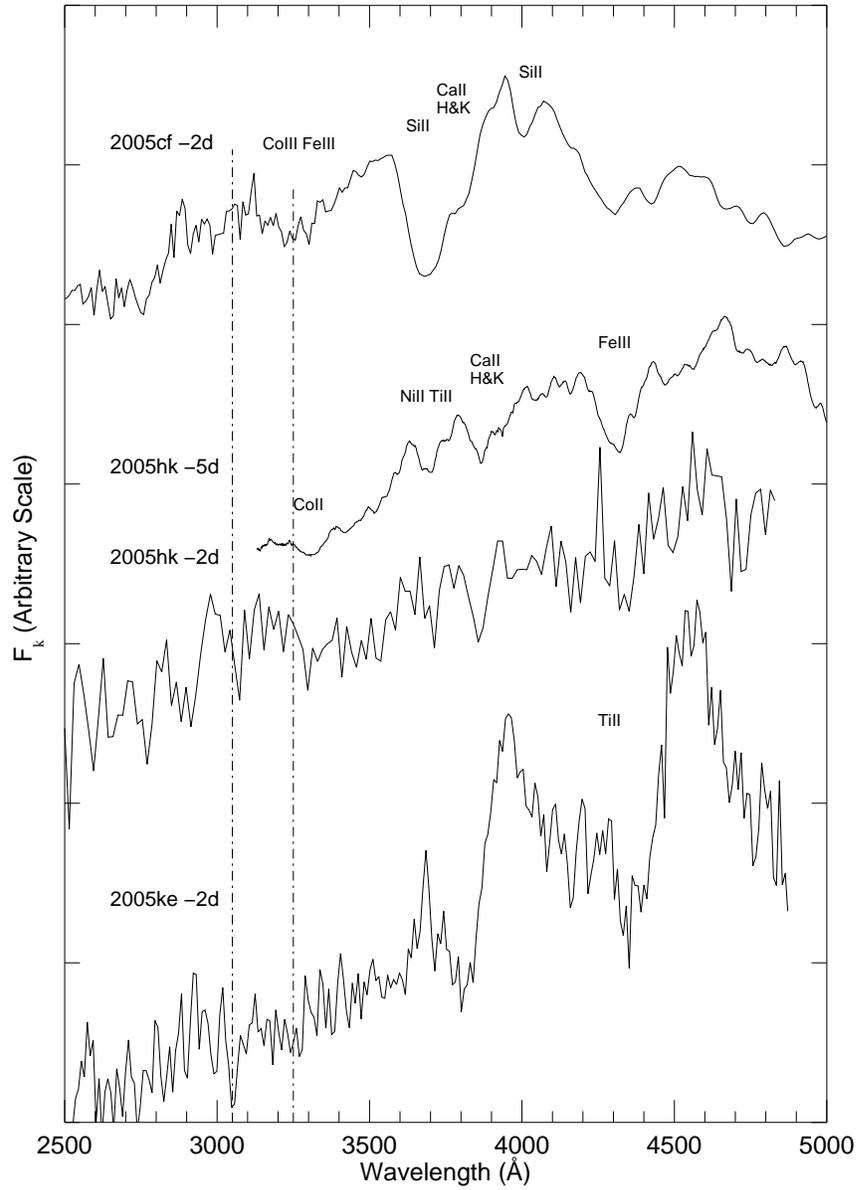}
\caption{Comparison of spectra for SNe 2005cf, 2005hk and 2005ke two days before maximum. SN~2005hk spectrum at $-5$ days was obtained with Keck/LRIS \citep{Chornock05hk}. Spectra are corrected for Galactic extinction \citep{Schlegel} and reported in the galaxy restframe, then scaled to the same magnitude and vertically displaced for clarity.  \label{conf_sub}}  
\end{figure}

Two peculiar SNe Ia are included in our sample: SN~2005ke, a sub-luminous 1991bg-like
SN, and SN~2005hk, a rare 2002cx-like SN.\\
The spectral evolution of SN~2005ke is shown in the lower panel of Fig.\  \ref{sn2005ke}.
All spectra were taken before maximum, but the wavelength calibration could only be checked 
for the V-grism spectrum at $-7$~d, by using the quasi-simultaneous
ground-based spectrum taken at VLT/FORS1 on Nov. 17 (courtesy of F.Patat), to correct for 
an apparent offset of the anchor point of 41\,\AA\ (Fig.\  \ref{sn2005ke}, upper panel). 
From the comparison we have recognized most
of the features and identified them on the basis of the similarities found
with the SN~1991bg-like SN~2005bl \citep{Tau05bl}.\\
SN~2005ke shows two broad absorption features, the CaII H\&K and Si II line at $\sim$3700\,\AA\ and the 
4200\,\AA\ TiII --- typical for this SN sub-class --- to which CII, SiII and MgII contribute.
Unlike normal SN~Ia, there is no clear evidence of 
the 3050\,\AA\ and 3250\,\AA\ minima in the UV region of the spectra,
as shown by the comparison (Fig.\ \ref{conf_sub}) with the spectrum
of SN~2005cf obtained at the same epoch.
The single UVOT spectrum of SN~2005hk  ($-2$~d) has a very poor S/N ($\sim$5) because of
the faintness of the object (${\rm mag}_B \sim16$, \citealt{phillips05hk}).
We also used the Keck/LRIS spectrum ($-5$~d, \citealt{Chornock05hk})
to recognize several narrow lines, indicative of low expansion
velocities: we note the $\lambda$4404 FeIII  and CaII\,H\&K, both at 
$\sim$6,000\,km\,s$^{-1}$, NiII plus TiII at about 3700\,\AA\ and CoII at 3300\,\AA.
 All these features were found previously only in SN~2002cx \citep{Li02cx}.

\subsubsection{SN~2006X}
\begin{figure}
\plotone{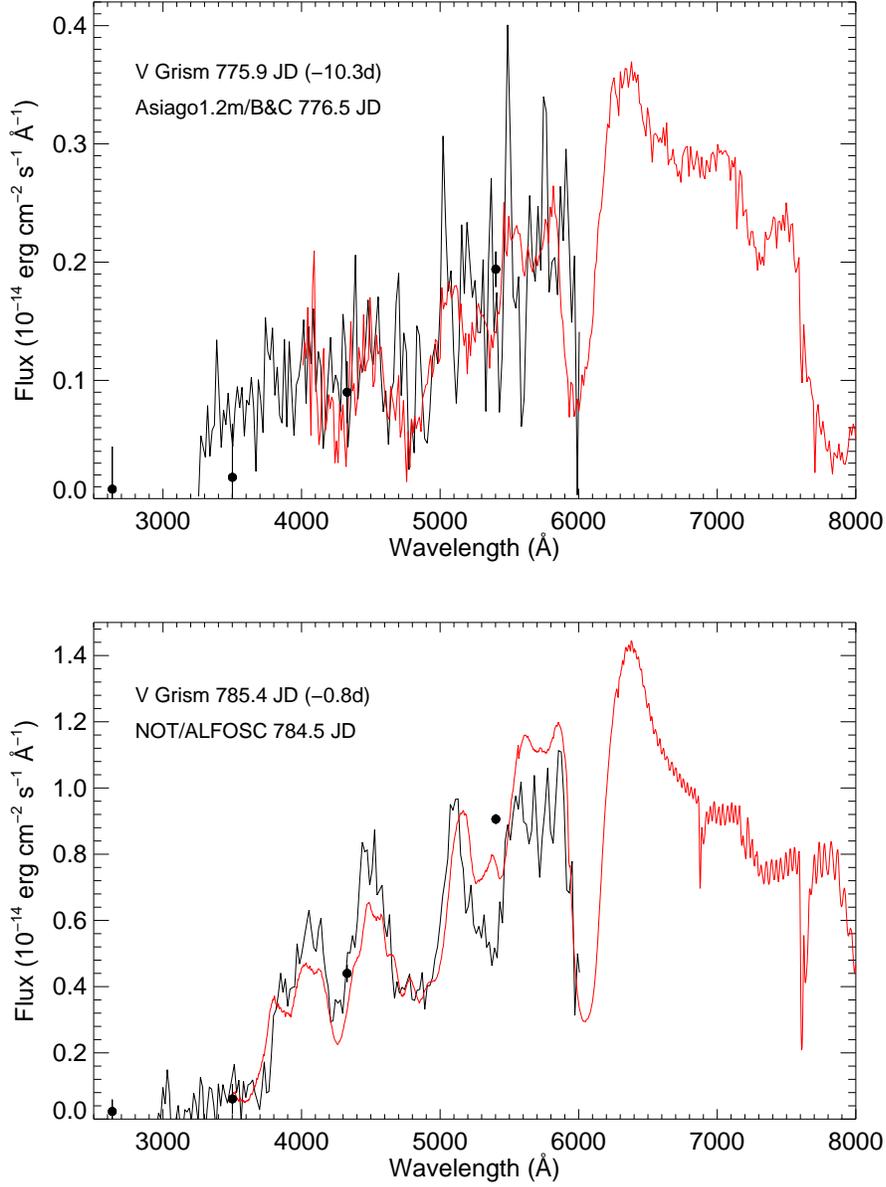}
\caption{ SN~2006X.  {\it Upper panel:} Superposition of the UVOT V-grism spectrum taken on day $-10.3$ (black line) and a quasi-simultaneous ground-based telescope (Asiago 1.2m/B\&C) spectrum (red line). Despite the low S/N, preventing the determination of the shift in wavelength, there is a good agreement between the SED of both spectra. {\it Lower panel:} Superposition of the UVOT V-grism spectrum taken on day $-0.8$ (black line) and a quasi-simultaneous ground-based telescope (NOT/ALFOSC ) spectrum (red line). A wavelength shift of 0\,\AA\ was found. Both UVOT spectra are scaled to match the \swift\ photometry (black points) and smoothed using a boxcar of 3 pixels ($\sim$18\,\AA). Ground-based spectra are courtesy of N. Elias-Rosa.  \label{conf_06x} }  
\end{figure}

The two pre-maximum V-grism spectra of SN~2006X are given in Fig.\ \ref{conf_06x},
together with ground-based data (Asiago 1.2m/B\&C and NOT/ALFOSC,
Elias-Rosa et al.\ 2009, in preparation). Because of the faintness of SN~2006X
and the short exposure time, the spectra remain noisy even after smoothing
with a boxcar of 18\,\AA\. For the first spectrum (upper panel in Fig.\
\ref{conf_06x}), it was not possible to compute from the cross-correlation with the ground-based
spectrum an unique solution for the wavelength
shift correction because of the strong noise, though good agreement with the SED is found. At shorter
wavelengths, possible contamination with the host galaxy might be present, as
suggested by a comparison with the UVOT photometry.  The second  
spectrum (lower panel) has better S/N, thus a wavelength offset of 0\,\AA\ 
has been found by cross-correlating it with the quasi-simultaneous ground-based
spectrum.  Numerous absorption lines of SiII 
($\lambda$ 4130, $\lambda$5051), SII ($\lambda$5468, $\lambda$5612), FeII 
($\lambda$4924) and FeIII at $\lambda$4404 blended with MgII $\lambda$4481 can be
recognized. SN~2006X shows a redder SED compared to normal SNe Ia, likely caused 
by strong circumstellar and interstellar absorption in the parent galaxy disk. 
  
\subsection {Core Collapse SNe}

Our sample of core collapse SNe includes two SNe II Plateau (SNe II-P 2005cs and 2006bp) and
one SN~Ib, the peculiar SN~2006jc. The \swift\ observations began soon 
after discovery of each of these targets and their spectral evolution was followed for 
as long as they remained detectable over the background. Preliminarily reduced UVOT
spectra were presented in \citet{Brown05cs} for SN~2005cs, \citet{Immler06bp} for
2006bp, and \citet{Immler06jc} for 2006jc. In this paper we present the spectra
obtained after applying the new extraction procedure, as described in \S\ \ref{datareduction}.

\subsubsection{Type II SNe 2005cs and 2006bp}
\begin{figure}
\plotone{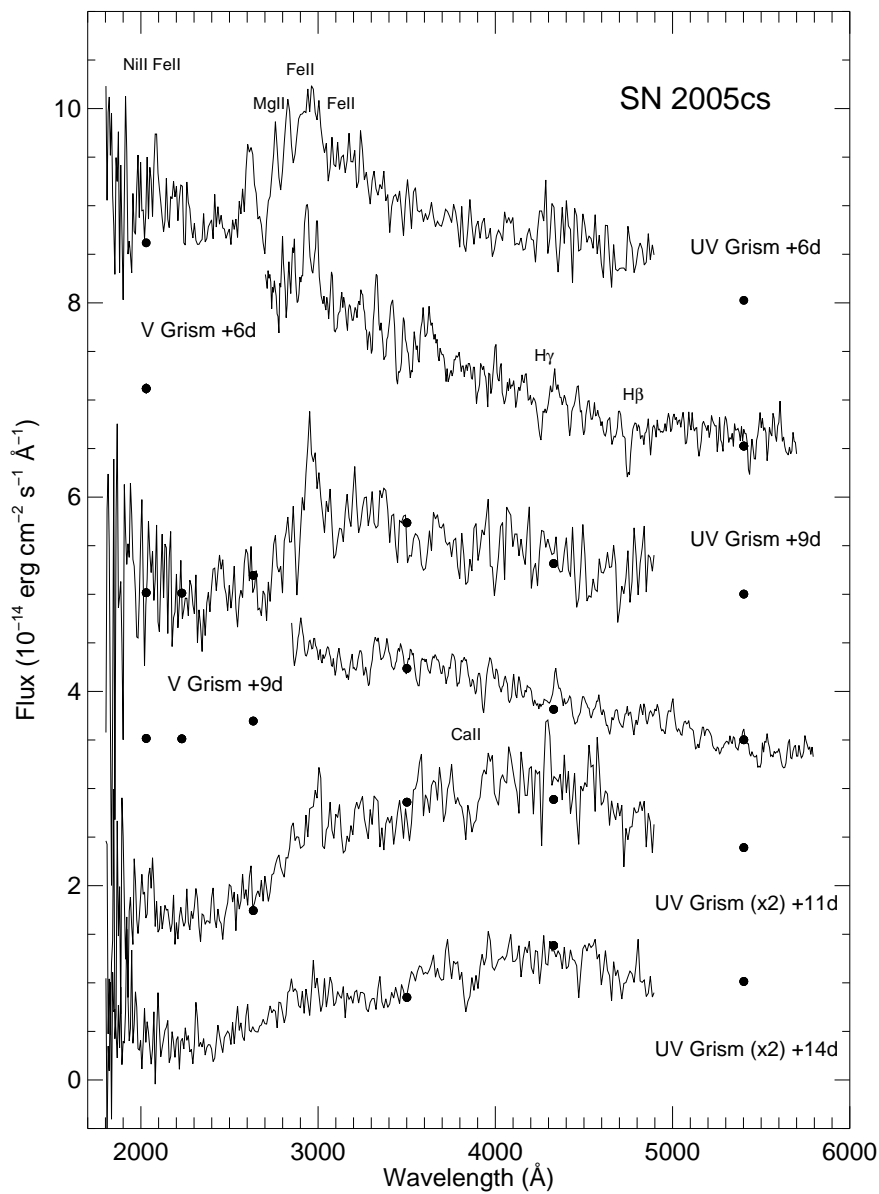}
\caption{\label{2005cs}Spectral evolution of SN~2005cs (Type II). All spectra are smoothed ($\sim$10\,\AA) and scaled to match the \swift\ photometry where available (black points) or  ground-based photometry \citep{pastorello05cs}. 
 No wavelength shifts are applied (see Fig.\ \ref{multi05cs}). The identifications of the strongest lines are marked. The spectra are vertically displaced by multiples of $1.5 \times 10^{-14}$ \flux.}
\end{figure}

The spectral evolution of SN~2005cs is shown in Fig.\ \ref{2005cs}. The extracted
spectra are smoothed using a boxcar of 10\,\AA\ and scaled to
match the quasi-simultaneous UVOT (\citealt{DessartUV}, \citealt{Brown05cs}) or
ground-based (Ekar 1.82m/AFOSC, \citealt{pastorello05cs}) photometry, where
available. Spectra taken after $+14$~d are not shown in Fig.\ \ref{2005cs}
because of poor S/N. In contrast to SN~Ia, SNe II are characterized by 
substantial UV emission, especially at early epochs. Since the UV-grism
covers a range from 2000\,\AA\ to 5000\,\AA, significant second-order
contamination is expected above 4000\,\AA.
The first UV-grism spectrum (2453555.2 JD) shows a
flux excess red-ward of $\sim4200$\,\AA. This excess amounts to $\approx 38$\% of the
simultaneous V-grism flux. Because of the rapid decrease of the 
UV emission, second-order contamination becomes rapidly negligible
(Fig.\ \ref{2005cs}). 
 
\begin{figure}
\plotone{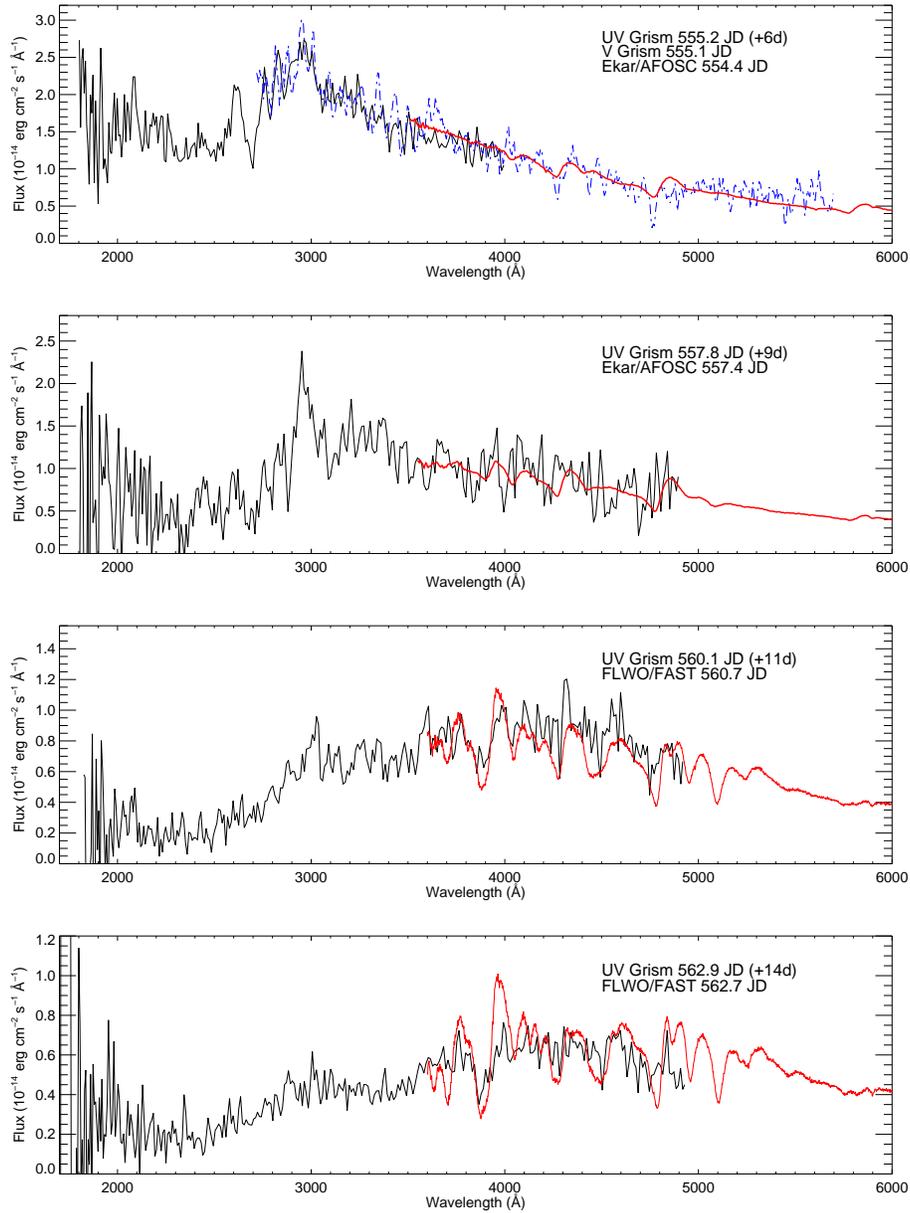}
\caption{Superposition of the UVOT UV-grism (solid black line), the V-grism (dashed blue line) and the quasi-simultaneous ground-based spectra (solid red line) of SN~2005cs, after wavelength offset correction (Tab.\ \ref{tabellashiftCC}).  Epochs (JD +2453000) and phases after the explosion are indicated. Ground-based spectra obtained at Asiago/Ekar have been taken from from \citealt{pastorello05cs}, and FLWO/FAST from \citealt{DessartUV}. \label{multi05cs} }  
\end{figure}

  Comparisons with ground-based spectra are shown in Fig.\ \ref{multi05cs}
and the derived wavelength offsets of the anchor point reported in Tab.\ \ref{tabellashiftCC}. 
In the top panel of Fig.\ \ref{multi05cs}, the
UVOT UV- and V-grism spectra taken on day $+6$ are compared with the
quasi-simultaneous Asiago spectrum \citep{pastorello05cs}. 
The part of the UV-grism spectrum with $\lambda > 4000$\,\AA\ has been removed as it was
contaminated by second-order light. An overall good agreement among the 3 spectra is found, and
Hydrogen Balmer lines (H${\beta}$ and H${\gamma}$) are easily recognizable. 
While the poor S/N of the UV-grism prevents us from estimating an unique solution for the wavelength 
shift correction, a 0\,\AA\ offset is derived for the V-grism spectrum. 
In Fig.\ \ref{confrontoII}, SN~2005cs spectrum shows significant resemblance to the HST
spectrum of SN~II-P~1999em (\citealt{Baron99em}, Fig.\ 2) although the latter 
corresponds to a somewhat later phase ($+11$~d).
In the UV, we can identify a broad absorption due to MgII
$\lambda$2798 and FeII absorption lines 
at about $2900$\,\AA\ and 3100\,\AA\ \citep{Dessart2005}.
The CaII H\&K doublet at about $3800$\,\AA\ is barely visible 
because of the poor S/N. It is evident only in the UVOT spectra taken
at $+11$ and $+14$~d (Fig.\ \ref{2005cs}). 
Similarly to SN~1999em \citep{Baron99em,Dessart2005,Dessart2006}, the spectrum of SN~2005cs is shaped at short wavelengths 
mainly by blends of lines of singly ionized iron-peak
elements.

  In Fig.\ \ref{multi05cs} (second panel from the top), we compare the UVOT UV-grism of
2005-07-06~UT with an optical spectrum obtained at Asiago/Ekar at the same
epoch. The agreement of the SEDs is excellent, but the S/N is again too low to estimate 
an unique wavelength shift correction from the cross-correlation. The second order effect seems 
already reduced at this late epoch.
The V-grism spectrum (not shown) has an even poorer S/N.
Fitting the UV-optical flux distribution with a black body at $\lambda >
3000$\,\AA\  to avoid the line-blanketing effect, we obtain color temperatures 
for the two early spectra of $\sim15,500$\,K and $\sim11,000$\,K, respectively,
with an uncertainty of 5\%. 
 
\begin{figure}
\plotone{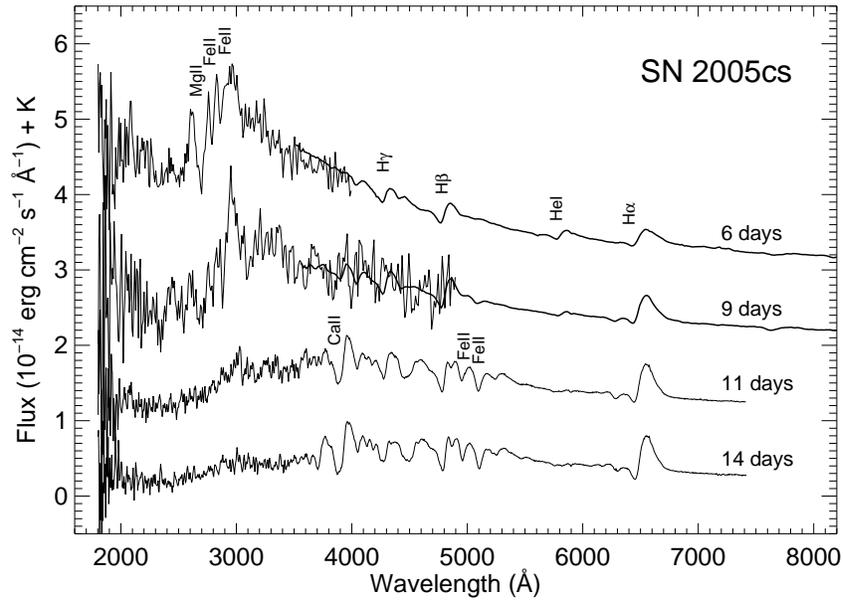}
\caption{  Evolution of the UV-optical spectral energy distribution of SN~2005cs. UVOT and ground-based spectra were
combined only where the anchor point wavelength correction was possible. Ordinates refer to the bottom spectrum while other spectra are vertically shifted by multiples of $1 \times 10^{-14}$ \flux. Identifications of the most prominent lines are also marked. \label{sn2005cs_comb} }
\end{figure}

 Finally, the last two epochs UV-grism spectra have been compared to the 
quasi-simultaneous optical ones taken at the FLWO 1.2m telescope (FAST spectrograph, 
\citealt{DessartUV}). We obtain  a wavelength shift of $-23$\,\AA\ and  $-33$\,\AA,
respectively (See Tab.\ \ref{tabellashiftCC}).\\
In Fig.\ \ref{sn2005cs_comb}, the time evolution of SN~2005cs spectral energy distribution in the range
1800--8000\,\AA\AA\ is presented, showing the combined UV-optical spectra where the correction of the
anchor point wavelength offset has been possible. During the early phases,
the UV flux is stronger than the optical
one, but the SN emission peak rapidly moves red-ward with time.  
The UV flux, produced during the SN shock breakout, is progressively blocked by
a forest of overlapping metal lines strengthening with time.\\
The synthetic spectra published by \citet{DessartUV} show a good agreement
with the observed UVOT spectra, here presented.  The models well reproduce the observed drop
of the continuum level blue-ward $\sim$2800\,\AA, due to the strong absorptions 
by FeII. An important contribution to the line blanketing is given by the TiII in the 3000\,\AA\ region 
and by NiII in the 2500\,\AA\ one.
We confirm the presence in the early spectra of broad absorptions centered approximately 
at 2350\,\AA\ and 2500\,\AA\ and the further continuum drop blue-ward 2000\,\AA\,
though a high noise characterizes this region. In addition, observed early spectra show  
a shallower MgII absorption at $\sim$2700\,\AA\ and  stronger FeII lines at  $\sim$2800\,\AA\ and
 $\sim$2900\,\AA\ than the corresponding models (Fig. 4c and Fig. 5b in \citealt{DessartUV}).
Such FeII lines appear only in synthetic spectra relative to later phases (Fig. 5c and Fig. 5e
in \citealt{DessartUV}), as well as the broad absorptions at $\sim$3100\,\AA\ and in the optical region. 

\begin{figure}
\plotone{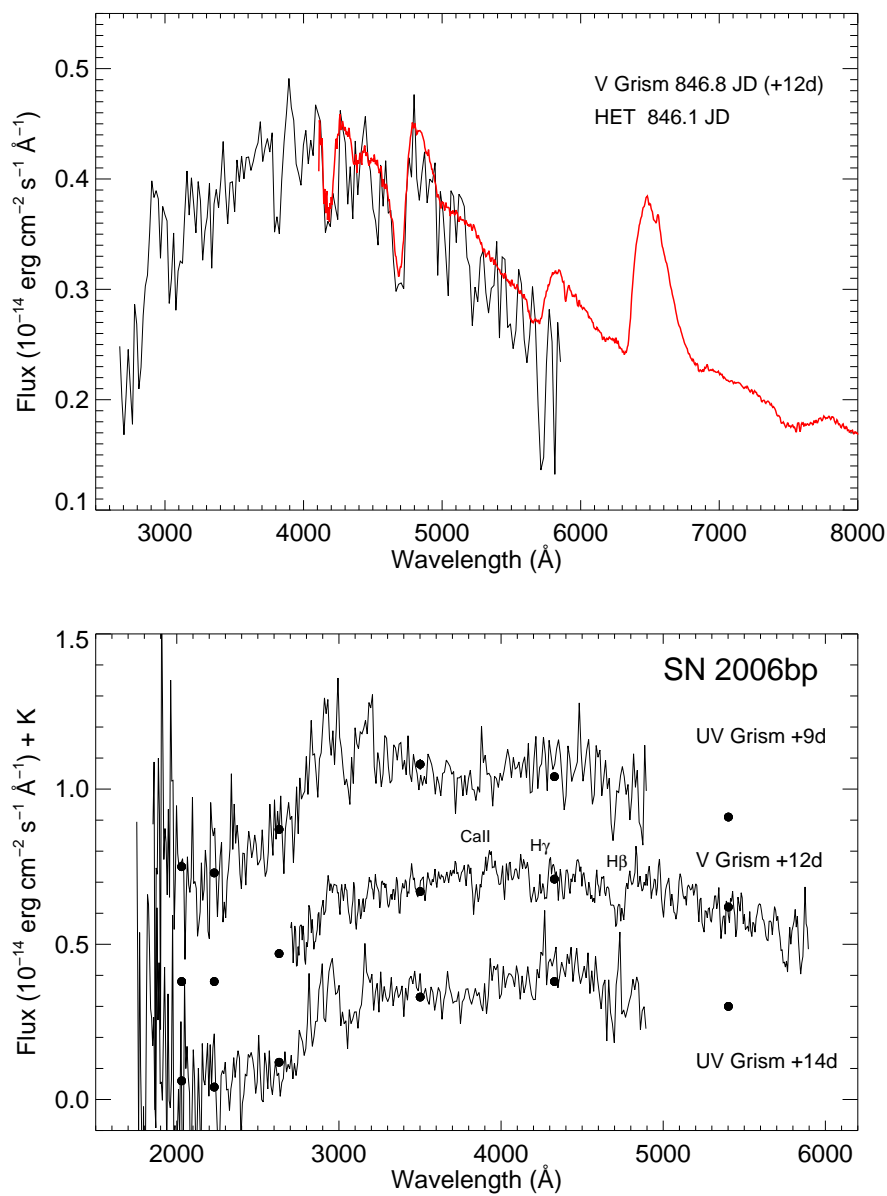}
\caption{SN 2006bp (Type II). {\it Upper Panel}:
Superposition of the UVOT V-grism (black line) and the quasi-simultaneous ground-based (HET) spectrum (red line, from \citealt{Quimby06bp}).   A wavelength offset of $-28$ \AA\ was measured. {\it Lower Panel}: Spectral evolution of SN 2006bp.  Spectra have been scaled to match Swift photometry (black points) and shifted by multiple of $3\times10^{-15}$\flux. No wavelength correction is applied.  \label{sn2006bp}}
\end{figure}

\begin{figure}
\plotone{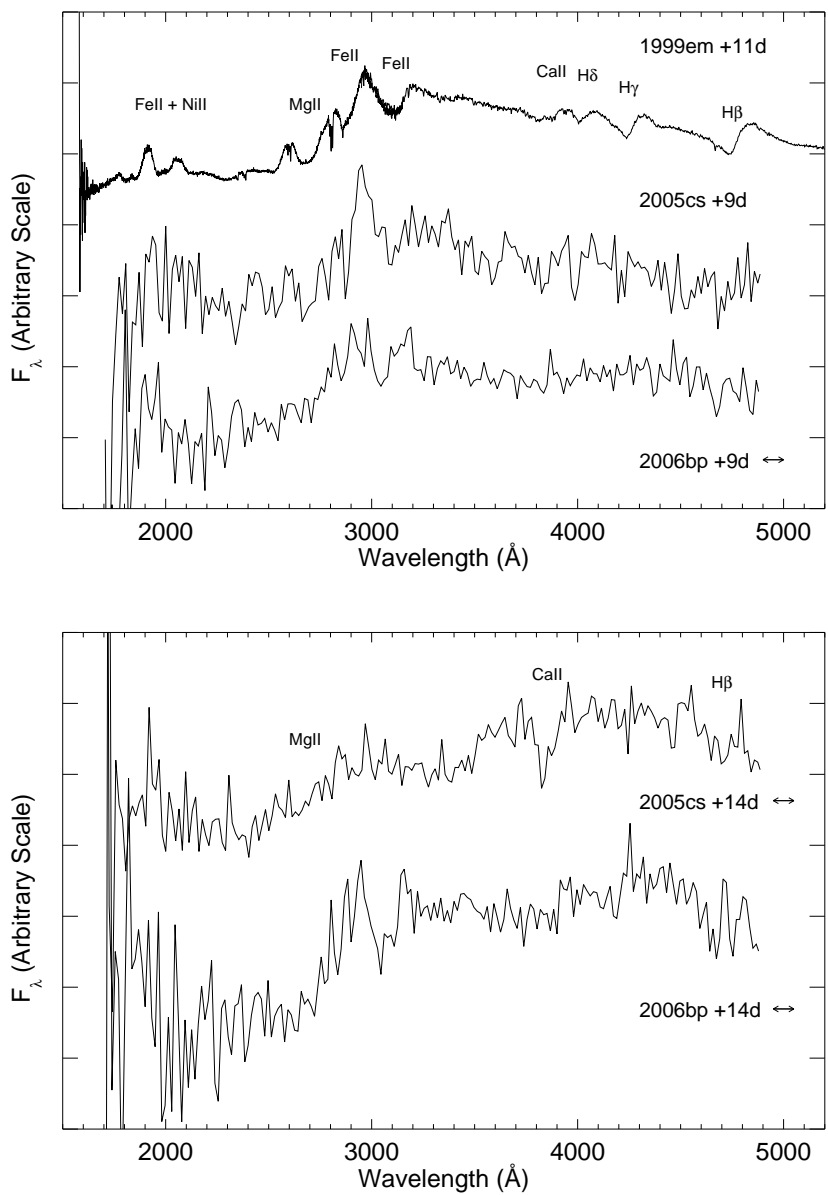}
\caption{{\it Upper panel}: Comparison of the UV spectra for SNe 2005cs, 2006bp and 1999em at about 9 days past explosion. {\it Lower panel}: SNe 2006bp and 2005cs spectra are compared at 2 weeks after the explosion. The epoch of explosion of SN~1999em comes from by \citet{E99em}. Spectra have been corrected for Galactic extinction \citep{Schlegel}, reported to the host galaxy rest-frame and smoothed with a boxcar of 5 pixels ($\sim$15\,\AA). 
Fluxes are scaled to similar magnitude and vertically displaced. A double-arrow indicates possible shifts in the wavelength calibration.\label{confrontoII}}  
\end{figure}

A single V-grism and seven UV-grism spectra were taken for SN~2006bp, 5 of which were
contaminated by a nearby field star. In Fig.\ \ref{sn2006bp} (lower panel), two UV
and the V-grism spectra are plotted. As described above, they have been flux calibrated
with the  simultaneous \swift\ photometry \citep{DessartUV} and smoothed
(boxcar of $\sim$10\,\AA).  Since the UV-grism spectrum has a very poor S/N, the wavelength calibration check 
was possible only for the V-grism one by cross-correlating it with a quasi-simultaneous ground-based spectrum 
taken at the HET (\citealt{Quimby06bp}). A shift of $\Delta\lambda=-28$\,\AA\ has been derived. 
The SED of the UVOT V-grism results in satisfactory agreement
with ground-based data (upper panel of Fig.\ \ref{sn2006bp}). 
The $3800$\,\AA\ CaII H \& K  is detectable only with the V-grism for the first epoch ($+9$~d).

The identification of the UV lines in SN~2006bp is facilitated by a comparison with other objects. In Fig.\ \ref{confrontoII}, we compare the UV-spectra of SNe 2006bp and 2005cs with those of SN~1999em at the same epoch \citep{Baron99em}.
We recognize FeII at $2900$\,\AA\, and $3100$\,\AA, and MgII at $2800$\,\AA\, \citep{Dessart2005} during the first epoch ($+9$~d). A blue-shift of the wavelength calibration scale is visible in SN~2006bp, taking SN~2005cs as a reference.  

Five days later (lower panel of Fig.\ \ref{confrontoII}), the continuum level decreases
and the spectra are redder because the ejecta cool down. Very few features can be identified.
The spectra of these two objects clearly show the need for considerably longer 
exposure times for future SN observations in order to improve the quality of the data and 
allow a more in-depth analysis and comparison with spectral models.

\subsubsection{Type Ib SN~2006jc}
\begin{figure}
\plotone{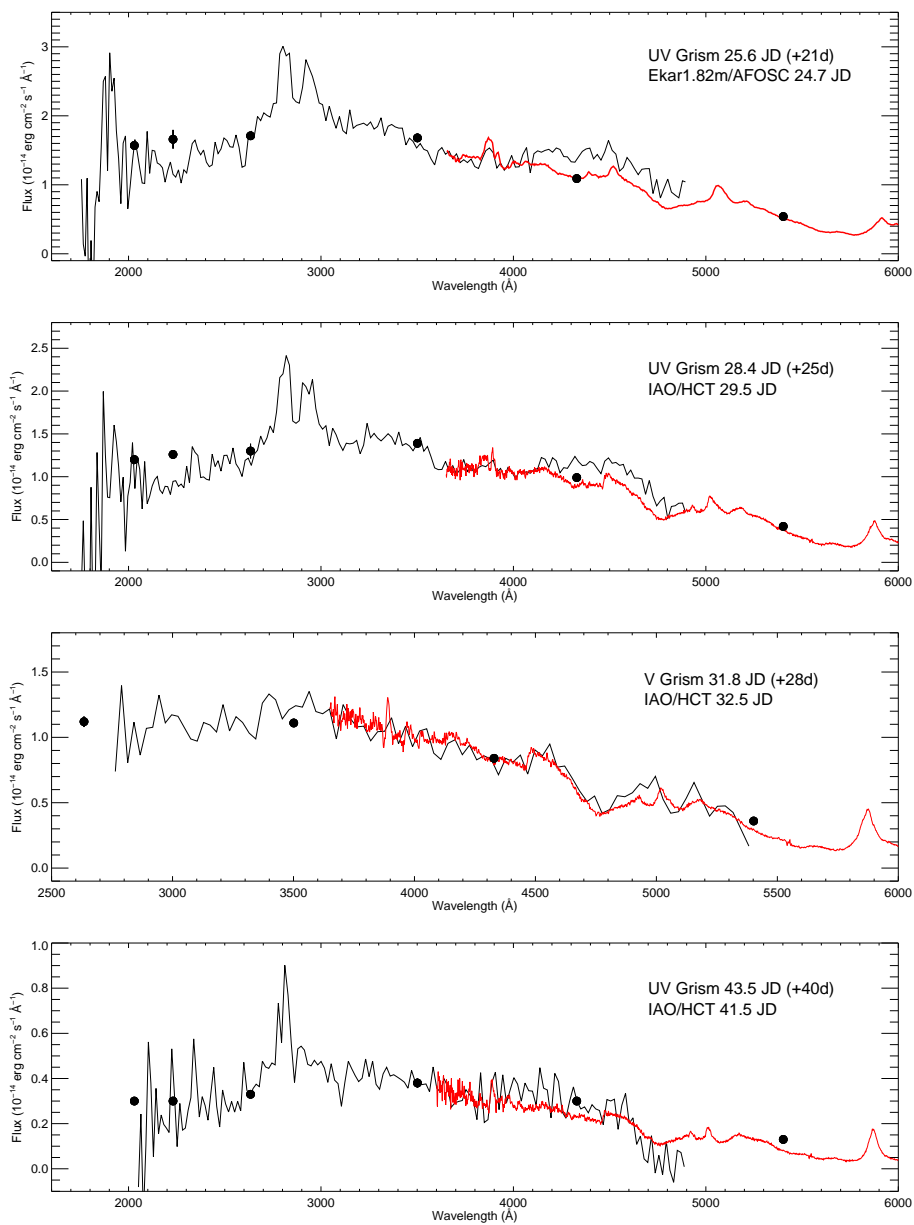}
\caption{  SN 2006jc. Superposition of the UVOT UV- and V-grism  spectra (solid black line) with the quasi-simultaneous ground-based optical one (solid red line) after wavelength offset correction (Tab.\ \ref{tabellashiftCC}). Spectra have been scaled to match Swift photometry (black points). Ground-based spectra obtained at Asiago/Ekar have been taken from from \citealt{pastorello06jc}, and IAO/HCT from \citealt{Anupama}. Epochs (JD +2454000) and phases after the explosion are indicated.\label{sn2006jc_over} }
\end{figure}

  For SN 2006jc, a cross-correlation was possible for all the  UVOT spectra (Fig.\ \ref{sn2006jc_over}),
comparing them with quasi-simultaneous ground-based optical spectra (Asiago/Ekar 
spectrum from \citealt{pastorello06jc} and IAO/HCT from \citealt{Anupama}).
Differently from the general constrain adopted along this paper (see \S\ 3.2), 
we compared the last epoch UV-grism spectrum (+40 days after explosion) with an optical
spectrum taken 2 days later. This choice was based on the slow evolution 
that characterizes SN~2006jc at this phase, as evident from the light curves 
(see \citealt{foley06jc} and \citealt{pastorello06jc}).

The resulting wavelength offsets  are negligible, as reported in Tab.\ \ref{tabellashiftCC}.
Similarly to the other SNe, the flux scale of each spectrum was matched to the UVOT photometry 
(Modjaz et al.\ 2009, in preparation), but at the earlier epochs ($+21$~d and $+28$~d) the comparison 
 with both the photometry and the ground-based spectra
reveals a flux excess red-ward of 4000\,\AA, caused by second-order contamination
(as discussed in the previous section). \\
 
\begin{figure}
\plotone{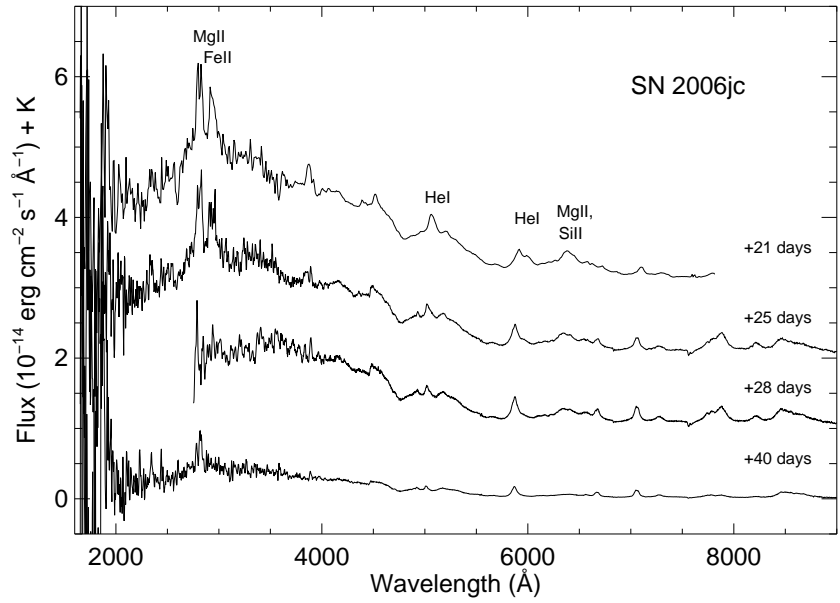}
\caption{ Evolution of the UV-optical combined spectrum of SN~2006jc. Ordinates refer to the bottom spectrum while other spectra are vertically shifted by multiples of $1 \times 10^{-14}$ \flux. The most prominent lines are also identified.\label{sn2006jc_comb} }
\end{figure}

The time evolution of the UV-optical combined spectrum (range 1800-9000\,\AA) of
SN~2006jc is shown in Fig.\ \ref{sn2006jc_comb}.  
A preliminary discussion of the spectral properties of SN~2006jc UV is reported in
\citet{Immler06jc}. Unlike SN~1983N \citep{Panagia1985}, the only SN~Ib 
extensively observed in the UV range by IUE, the spectral features of SN~2006jc do not 
resemble those of a Type Ia SN.
Absorption features are recognizable at $\sim$2400\,\AA\
and $\sim$2600\,\AA, but  detailed fitting with a spectral model is needed to identify the
elements that cause them. 
SN~2006jc shows two strong and broad emission lines at 2800\,\AA\ and 2950\,\AA,
 probably MgII and FeII, respectively. 
The appearance of these features in emission at early epochs is rather unusual.
The MgII $\lambda$2800 emission line has been observed in the spectra of 
SN~IIL 1979C \citep{Panagia1980} and SN~IIn\,1998S (\citealt{Panagia2003}, \citealt{Fransson93J98S}). 
In these SNe  the emission line became stronger relative to the continuum 
after maximum, and broader. Such an evolution was explained 
as the direct result of the interaction of the SN ejecta with a circumstellar
medium. MgII emission lines form mainly in layers that are close to the SN 
photosphere, but distinctly separated from it in space, velocity and excitation conditions
 (\citealt{Panagia1980}, \citealt{Fransson1984}). 
Such a UV emitting shell may consist of gas originally ejected by the stellar
progenitor as a stellar wind or during an episodic mass ejection.
MgII emission is more commonly observed in late phase spectra, e.g., for SN~1995N \citep{Fransson2002}, of the previously quoted SN~1979C \citep{Fesen79c}, or of the SN~1993J 
and SN~1998S \citep{Fransson93J98S}, and is interpreted as interaction
with pre-existing CSM shell. The X-ray and UV photons emitted in the propagation of the radiative reverse shock into the supernova ejecta, highly ionize the unshocked material (with resulting CIII-IV, NIII-IV and OIII-IV lines in the far UV). At the same time, the reprocessed radiation emerges mostly as emission from neutral and singly ionized ions (such as HI lines, MgII and FeII) 
because of the high density of the region between the reverse shock and the contact discontinuity
\citep{Fransson93J98S}. 
 
Observations at different wavelength ranges give independent
evidence for CSM interaction. X-ray detections were obtained with {\it Swift}/XRT
and Chandra at seven epochs, which showed that the flux increased by a factor of
5 up to about $120$ days after the explosion. This unusual X-ray rise and subsequent decline has
never been observed for any other SN and could be explained as shock-heating of the
previously ejected progenitor shell by the SN blast wave \citep{Immler06jc}.  
Indeed, SN~2006jc has been associated with a luminous outburst 
discovered two years earlier (\citealt{nakano06jc}, \citealt{pastorello06jc}),
which may indicate that the progenitor was a Wolf-Rayet star with activity similar to a
Luminous Blue Variable (LBV).  This scenario is also consistent with the strong and
narrow HeI emission lines observed in the optical and IR (\citealt{pastorello06jc};
\citealt{foley06jc}).  
 
\section{Conclusions and Outlook}

We have presented all \swift\ archival of UV/optical spectroscopic data of SNe
obtained during the first two years of observations (2005--2006). The sample includes  29  UV
grism and 22 V grism spectra for a total of 9 SNe (6 SNe Ia, 2 SNe II and
1 SN Ib). In SNe Ia the UV spectrum is thought to originate predominantly from reverse
fluorescence in iron group ions in the outermost layers that converts photons
from red to blue wavelengths \citep{Mazzali2000}, thus the emerging UV spectrum
changes with the metal content of the outermost layers of the ejecta (line
blanketing effect) and its ionization (backwarming effect) (\citealt{Lentz2000},
\citealt{sauer08}). We confirmed that the broad absorptions at around $3050$\,\AA\
 and $3250$\,\AA\ [due to Fe, Co, Cr and Ti lines; \citet{Kirshner92A}; \citet{sauer08}],
 along with the 2800\,\AA\ MgII, are the main features characterizing the UV spectrum of SNe Ia. 
A comparison of the \swift\ SN~Ia spectra with those of SNe 1992A and 1990N shows that SNe Ia 
with similar optical properties may have different properties in the UV.

SN~2006jc is the only Type Ib/c SN in our spectroscopic sample, although it is peculiar in
many aspects. Unlike the previous UV-detected SN~Ib 1983N \citep{Panagia2003},
the UV spectra of SN~2006jc do not show features similar to SNe Ia.
Two strong and broad emission features are present in the range  2800--3000\,\AA.
These lines are unusual at early epochs and are likely due to MgII.
Resembling the strongly interacting SNe 1979C \citep{Panagia1980} and 1998S 
(\citealt{Panagia2003}, \citealt{Fransson93J98S}), the emission can be interpreted as a 
signature of the interaction between the SN ejecta and material originally ejected by the progenitor.

Most of the UV flux of SNe II is emitted during the shock breakout.
In some cases, a strong UV excess is also seen, together with highly ionized
species lines, as a consequence of the inverse Compton scattering by energetic, thermal electrons 
($T \sim10^9$~K), produced when the SN ejecta interacts with the pre-existing
CSM \citep{Fransson1984}.
Similar to SN~1999em \citep{Baron99em}, the UV spectra of both  Type II-P 
SNe 2005cs and 2006bp are shaped by blends of singly ionized iron-peak elements
lines (FeII, NiII)  and MgII.
The presence of numerous resonance lines yields a line blanketing effect that
increases with decreasing temperature.  
 SNe II-P constitute a homogeneous   class in the UV wavelength range.
This confirms the recent claims by \citet{Gal-Yam}, who found, based on the GALEX UV spectrum of SN~2005ay,  
a remarkable similarity among the UV spectral properties of
these SNe. This open new perspectives for their use as cosmological probes.
A deep discussion on the UV homogeneity of Type II-P SNe
and the comparison with the UV properties of other Type II classes is presented in Bufano et al. (2009, in prep.).

Our analysis is limited by the low S/N of the data as a result of short exposure times.
However, the data presented here highlight the large and unique scientific potential of \swift\ 
grism observations of SNe and justify the continuation of the ongoing \swift\ program to monitor SNe of all types
with a sufficiently large investment in exposure time.  

\begin{figure}
\plotone{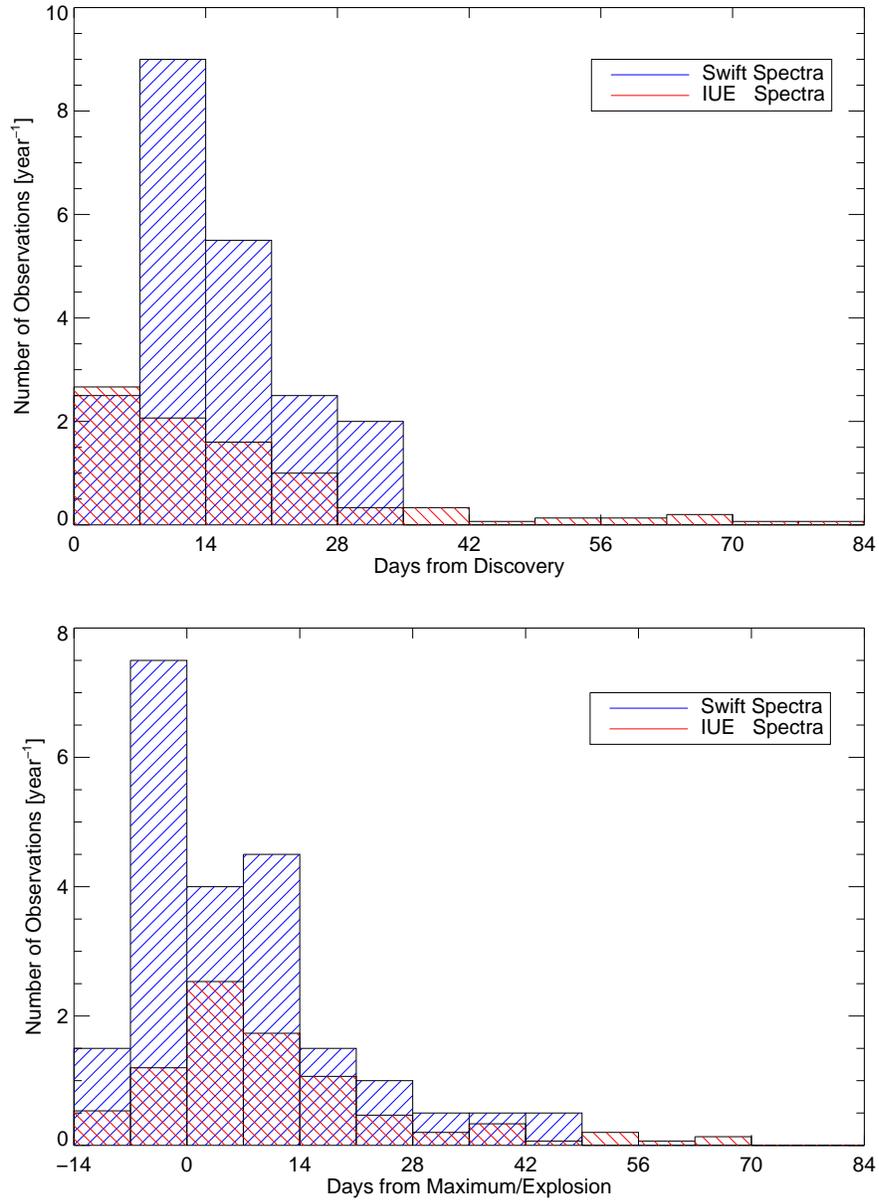}
\caption{\label{hist} A histogram of the total number of UV spectra of SNe per year and per phase bin. Phases are relative to the day of the discovery ({\it upper panel}) and to the day of the $B$-maximum light for SNe Ia or explosion for CC SNe ({\it lower panel}). SN observations were collected during a 15-years long campaign for IUE (red) and 2-years long campaign for \swift\ (blue).  }
\end{figure}

Currently (and for years to come) \swift\ is the only instrument capable of monitoring the
rapid evolution of SN UV emission. By using its flexible scheduling capabilities, 
we have obtained prompt observations and carried out intensive follow-up campaigns. 
For SN~2005cf, for example, the best SN Ia spectroscopic UV observations ever have been obtained  
 that include 7 UV spectra before maximum. This  improves   the
exceptional performance of IUE (6 pre-maximum spectra of SN~1990N). 
In Fig.\ \ref{hist} we compare the
first 2-years of \swift\  spectroscopic visits  to the 15-year {\sl IUE} activity on SNe.
 Our results are encouraging, especially considering the small collecting
area of \swift\ UVOT. The large number of observations that \swift\ already collected stands out, as
a result of its unsurpassed capabilities:  it has a faster and   flexible
response time ($\leq$1 day)  making feasible systematic follow-up   of SNe starting from
early phases, when the SN evolves faster, the UV emission is stronger and the photosphere 
is located in the outermost layers that still retain information on the SN progenitors.
However, a more aggressive strategy for brighter SNe, 
especially closer to the discovery, when  \swift\ remains slightly lower than  {\sl IUE}, is recommended.  
Future observations will probe further the diversity of SNe of all types, their environments, 
constrain theoretical stellar models and test their possible cosmological evolution, 
fundamental for using SNe as distance indicators.

\acknowledgments
This work made use of public data from the \swift\ data 
archive and the NASA/IPAC Extragalactic Database.
We thank F.~Patat and N.~Elias-Rosa for providing their optical spectra 
of SN~2005ke and SN~2006X, respectively, and M.~Page for his
useful feedback.
FB acknowledges financial support from the NASA \swift\ project and from 
ASI/INAF (grant n. I/088/06/0).
SB, EC and MT are supported by the Italian Ministry of Education
via the PRIN 2006 n.2006022731 002.
The spectrum of SN2005ke has been obtained with ESO Telescopes at Paranal 
Observatory under programme ID 076.D-0178(A).
Based also on observations made with the NASA/ESA Hubble Space Telescope, obtained from the data archive at the Space Telescope Institute. STScI is operated by the association of Universities for Research in Astronomy, Inc. under the NASA contract NAS 5-26555.

 

\clearpage
 
\begin{deluxetable}{cccccc}
\tabletypesize{\scriptsize}
\tablecaption{Main SN parameters \label{SNsample} }
\tablewidth{0pt}
\tablehead{SN &Type& Coordinates [J2000.0]&Host Galaxy&Offset [arcsec]&{\it z}\\
(1)&(2)&(3)&(4)&(5)&(6)}
\startdata
SN~2005am&Ia&$09^h16^m12^s.47\quad -16^{\circ}18'01''.0$&NGC 2811      &17E\quad31N&0.007\\
SN~2005cf&Ia&$15^h21^m32^s.21\quad +07^{\circ}24'47''.5$&MCG-01-39-003 &15W\quad123N&0.006\\
SN~2005cs&II&$13^h29^m52^s.85\quad +47^{\circ}10'36''.3$&M51           &15W\quad67S&0.002\\
SN~2005df&Ia&$04^h17^m37^s.85\quad -62^{\circ}46'09''.5$&NGC 1559      &15E\quad40N&0.004\\
SN~2005hk&Ia&$00^h27^m50^s.90\quad -01^{\circ}11'52''.5$&UGC 272       &17E\quad6N&0.013\\
SN~2005ke&Ia&$03^h35^m04^s.35\quad -24^{\circ}56'38''.8$&NGC 1371      &40E\quad40S&0.005\\
SN~2006X &Ia&$12^h22^m53^s.99\quad +15^{\circ}48'33''.1$&M100          &12W\quad48S&0.005\\
SN~2006bp&II&$11^h53^m55^s.70\quad +52^{\circ}21'10''.4$&NGC 3953      &62E\quad93N&0.003\\ 
SN~2006jc&Ib/pec&$09^h17^m20^s.81\quad+41^{\circ}54'32''.9$&UGC 4904       &11W\quad7S&0.005\\
\enddata
\tablecomments{(1) SN name; (2) SN Type; (3) R.A. and Dec.; (4) host galaxy name; 
(5) SN positional offset with respect to the nucleus of the host galaxy; (6) red-shift of the host galaxy, from NASA/IPAC Extragalactic Database (NED) and references therein.}

\end{deluxetable}

\begin{deluxetable}{cccccc}
\tabletypesize{\scriptsize}
\tablecaption{UVOT grism observations. \label{observation}}
\tablewidth{0pt}
\tablehead{SN (Type)& Grism &  Date & JD &Exposure Time & Epoch\\ 
(1)&(2)&(3)&(4)&(5)&(6)}
\startdata
SN~2005am (Ia) &UV& 2005-03-08&2453437.5& 1657.8&-1\tablenotemark{*}\\
&V& 2005-03-08&2453438.2&1784.1& 0\\
&UV&2005-03-09& 2453439.1&2232.4&+1\\
&V&2005-03-10&2453440.1&2290.3&+2\\
&V&2005-03-15&2453445.4&1800.0&+7\\
&V&2005-03-17&2453447.3&1799.4&+9\\
&UV&2005-03-18&2453448.4&1533.5&+10\tablenotemark{*}\\ 
&UV& 2005-03-22&2453451.6&1777.6&+14\tablenotemark{*}\\
&V&2005-03-23&2453452.6&1810.4&+15\\
&UV& 2005-03-24&2453454.0&1441.0&+16\tablenotemark{*}\\
SN~2005cf (Ia)& UV& 2005-06-04&  2453526.2& 1577.1&$-$7.8\\ 
&V&2005-06-04&2453526.3&1391.5&$-$7.7\\
&UV& 2005-06-05&  2453527.2& 1626.9& $-$6.8\\ 
&V&2005-06-05&2453527.3&993.4 & $-$6.7\\
& UV &2005-06-06& 2453528.2& 1385.6&$-$5.8 \\
&V&2005-06-06&2453528.3&1371.4&$-$5.7\\
&V&2005-06-07&2453529.3&55.3& $-$4.7\\
& UV&2005-06-08&2453530.3& 1438.3&$-$3.7 \\ 
&UV&2005-06-09&2453531.0& 531.7& $-$3.0\\
&UV&2005-06-10&  2453532.2&460.9 & $-$1.8\\ 
&V&2005-06-10&2453532.3&1544.0&$-$1.7\\
&  UV&2005-06-11&2453533.1& 1679.7&$-$0.9 \\ 
&V& 2005-06-11&2453533.2&1580.5&$-$0.8\\
&V&2005-06-15&2453536.6& 1511.4&+2.6\\
&UV&2005-06-16&   2453537.8& 1507.6&+3.8 \\
&V&2005-06-16&2453538.4& 1505.6&+4.4\\ 
&UV&2005-06-17&   2453538.8& 1342.4& +4.8\\
&V&2005-06-17&2453539.2& 1647.4&+5.2\\
&UV&2005-06-20&   2453542.2& 1663.1& +8.2\\
&V&2005-06-20& 2453542.3& 1564.7&+8.3\\
&UV&2005-06-26& 2453547.8& 1665.1&+13.8\\ 
&UV&2005-06-29&   2453550.6& 1767.2&+16.6\\

SN~2005cs (IIP)& V&2005-07-03&2453555.1 &2045.3&+6\\
&UV&2005-07-03&2453555.2& 2041.4& +6\\
&V&2005-07-06&2453557.7&2104.4& +9\\ 
&UV&2005-07-06&  2453557.8&2102.0&+9 \\ 
&UV&2005-07-08& 2453560.1&1985.9 &+11\\ 
&UV&2005-07-11&    2453562.9& 2042.8&+14\\ 
&UV&2005-07-13&    2453564.6& 1760.1&+16\tablenotemark{**}\\ 
&UV&2005-07-19&   2453570.7& 1924.4&+22\tablenotemark{**}\\ 
SN~2005df (Ia)& UV&2005-08-11&2453593.5& 1608.9 &$-$7\\
 &UV&2005-08-14& 2453596.6& 976.9&$-$4\\
  &UV&2005-08-17& 2453600.4  &1986.3&$-$1\\
 &UV& 2005-08-21&2453604.1&415.4&+3\\
SN~2005hk (Ia)& UV&2005-11-08& 2453682.7&2014.9&$-$2.4\\ 
SN~2005ke (Ia)& UV&2005-11-15& 2453690.0& 1836.0&$-$9\tablenotemark{**}\\
&V& 2005-11-17&2453692.3& 478.8&$-$7\\ 
 &UV& 2005-11-20&2453694.9 &1633.4&$-$4\\
 &UV& 2005-11-22&2453696.9 &3068.6&$-$2\\
 SN~2006X (Ia) & V & 2006-02-09 & 2453775.9&1263.5 & $-$10.3\\
 & V & 2006-02-18 & 2453785.4&2740.0 & $-$0.8\\
 SN~2006bp (IIP)&UV& 2006-04-12&2453837.6& 662.1&+3\tablenotemark{*}\\
&UV&2006-04-14&2453839.8& 607.9&+5\tablenotemark{*}\\
&UV&2006-04-14&2453840.2& 768.4&+5\tablenotemark{*}\\
&UV& 2006-04-16&2453841.8&2479.7&+7\tablenotemark{*}\\
&UV& 2006-04-16&2453842.2& 1264.47 &+7\tablenotemark{*}\\
& UV&2006-04-18& 2453843.3&2430.5&+9\\
&V&2006-04-21&2453846.8&2697.6&+12 \\ 
&UV&2006-04-23& 2453848.6& 3202.9&+14\\ 
SN~2006jc (Ib)& UV&2005-10-17& 2454025.6& 2024.1&+21\\
&UV& 2006-10-20&2454028.4&2201.0 &+25\\
&V&  2006-10-23&2454031.8& 1492.2&+28\\ 
&UV& 2006-11-04&2454043.5& 2909.7&+40\\
 &V& 2006-11-04&2454043.6& 1910.0& +40\tablenotemark{*}\\ 
 &V& 2006-11-19&2454059.0& 3820.0& +46\tablenotemark{*}\\ 
\enddata

\tablenotetext{*}{ Spectrum contaminated by field star.}
\tablenotetext{**}{ Very low S/N spectrum (average S/N $<$5).}
\tablecomments{(1) Name of the SN and Type; (2) UVOT grism used; (3) observation date (yy-mm-dd);
 (4) start of the UV grism observation in Julian Days; (5) exposure time in unit of seconds;
(6) days after the $B$-band maximum light for SNe Ia or after explosion day for CC SNe 
(Type Ib/c and Type II)}
\end{deluxetable}

\begin{deluxetable}{ccccr}
\tabletypesize{\scriptsize}
\tablecaption{ Anchor point offsets of the wavelength calibration for SN~2005cf. \label{tabellashift}}
\tablewidth{0pt}
\tablehead{\multicolumn{2}{c}{\swift\  Spec.}& \multicolumn{2}{c}{Ground-Based Spec.$^{a}$} &  Offset \\
Grism & JD [days]&Teles./Instrum.& JD [days]&[\AA]}   
\startdata
UV &2453526.2&CA3.5m/PMAS&2453525.4& 0  \\
V &2453526.3&CA3.5m/PMAS&2453525.4& 0   \\
UV &2453527.2&TNG/DOLORES&2453527.6& 8  \\
V &2453527.3&TNG/DOLORES&2453527.6&$-$22  \\
UV &2453528.2&NTT/EMMI&2453528.7&24 \\
V &2453528.3&NTT/EMMI&2453528.7&13 \\
UV &2453530.3& NOT/ALFOSC&2453530.4&12 \\
UV &2453531.0&NOT/ALFOSC&2453530.4& 16 \\
UV &2453532.2&NOT/ALFOSC&2453531.5&6 \\
V&2453532.3&NOT/ALFOSC&2453531.5&12 \\
UV &2453533.1&NOT/ALFOSC&2453533.0&19 \\
V &2453533.2&NOT/ALFOSC&2453533.0&$-$27 \\
V &2453538.4&CA2.2m/CAFOS&2453538.4&$-$56\\
UV &2453538.8&CA2.2m/CAFOS&2453538.4&$-$8 \\
V &2453539.2&CA2.2m/CAFOS&2453538.4&$-$66 \\
\enddata
\tablenotetext{a}{Data from \citet{Garavini05cf}.}
\end{deluxetable}

 \begin{deluxetable}{ccccr}
\tabletypesize{\scriptsize}
\tablecaption{ Anchor point offsets of the wavelength calibration for the CC SNe sample. \label{tabellashiftCC}}
\tablewidth{0pt}
\tablehead{\multicolumn{2}{c}{\swift\  Spec.}& \multicolumn{2}{c}{Ground-Based Spec.$^{a}$} &  Offset \\
Grism & JD [days]&Teles./Instrum.& JD [days]&[\AA]}  
\startdata
&&SN 2005cs&&\\
\hline
V &2453555.1&Ekar/AFOSC$^{1}$&2453554.4 &  0 \\
UV& 2453555.2&Ekar/AFOSC$^{1}$& 2453554.4& --$^{*}$\\
V &2453557.7&Ekar/AFOSC$^{1}$& 2453557.4& --$^{*}$  \\
UV&2453557.8&Ekar/AFOSC$^{1}$& 2453557.4& --$^{*}$ \\
UV&2453560.1&FLWO/FAST$^{2}$&2453560.7 & -23\\
UV&2453562.9&FLWO/FAST$^{2}$& 2453562.7& -33\\
\hline\noalign{\smallskip} 
&&SN 2006bp&&\\
\hline
UV& 2453843.3 &McDonald Obs/HET$^{3}$ &2453844.1& --$^{*}$\\
V & 2453846.8 &McDonald Obs/HET$^{3}$ &2453846.1  &-28\\
UV& 2453848.6 &FLWO/FAST$^{2}$& 2453849.7&--$^{*}$\\
\hline\noalign{\smallskip} 
&&SN 2006jc&&\\
\hline
UV &2454025.6&Ekar/AFOSC$^{4}$&2454024.7& 0\\
UV &2454028.4& IAO/HCT$^{5}$&2454029.49& 0\\
V  &2454031.8&IAO/HCT$^{5}$&2454032.46& 0\\
UV &2454043.5&IAO/HCT$^{5}$&2454041.48& 4\\
\enddata
\tablenotetext{a}{Data from: (1) \citet{pastorello05cs}; (2) \citet{DessartUV};
(3) \citet{Quimby06bp}; (4) \citet{pastorello06jc}; (5) \citet{Anupama}.}
\end{deluxetable}
\tablenotetext{*}{Very low S/N spectrum. No unique solution found from the automatic cross-correlation.}

\epsscale{.75}

\clearpage 

\clearpage


\begin{thebibliography}{}
\bibitem[Anupama et al.(2009)]{Anupama} Anupama, G.~C., Sahu, 
D.~K., Gurugubelli, U.~K., Prabhu, T.~P., Tominaga, N., Tanaka, M., 
\& Nomoto, K.\ 2009, \mnras, 392, 894 


\bibitem[Baron et al.(2000)]{Baron99em} Baron, E., et al.\ 2000, \apj, 545, 444 
\bibitem[Barthelmy et al.(2005)]{barth04} Barthelmy, S.~D., et 
al.\ 2005, Space Science Reviews, 120, 143 
\bibitem[Baek et al.(2005)]{Puckett05ke} M. Baek, M., Prasad, R.~R. \&  Li, W.  \ 2005, IAU Circ. 8630
\bibitem[Benetti et al.(2004)]{Benetti02bo} Benetti, S., et al.\ 2004, \mnras, 348, 261 
\bibitem[Benetti et al.(2006)]{benetti06jc}Benetti, S., et al. \ 2006, CBET, 674 
\bibitem[Benvenuti et al.(1982)]{Benvenuti82} Benvenuti, P., Sanz Fernandez de Cordoba, L., Wamsteker, W., Macchetto, F., Palumbo, G.~C., Panagia, N., \& Battrick, B.\ 1982, NASA STI/Recon Technical Report N, 84, 25563 

\bibitem[Bianchi \& The GALEX Team(2000)]{BianchiGALEX} Bianchi, L., \& The GALEX Team 2000, Memorie della Societa Astronomica Italiana, 71, 1117 
\bibitem[Blinnikov \& Bartunov(1993)]{Blinnikov1993} Blinnikov, S.~I., \& Bartunov, O.~S.\ 1993,  \aap, 273, 106 
\bibitem[Blinnikov et al.(1998)]{Blinnikov1998} Blinnikov, S.~I., Eastman, R., Bartunov, O.~S., Popolitov, V.~A., \& Woosley, S.~E.\ 1998, \apj, 496, 454 
\bibitem[Blinnikov \& Sorokina(2000)]{Blinnikov2000} Blinnikov, S.~I., \& Sorokina, E.~I.\ 2000,  \aap, 356, L30 
 \bibitem[Branch et al.(1985)]{Branch85model} Branch, D., Doggett, J.~B., Nomoto, K., \& Thielemann, F.-K.\ 1985, \apj, 294, 619 
\bibitem[Branch \& Venkatakrishna(1986)]{BranchUVspectra} Branch, D., \& Venkatakrishna, K.~L.\ 1986, \apjl, 306, L21
\bibitem[Branch(1987)]{Branch1987A} Branch, D.\ 1987, \apjl, 320, L23 
\bibitem[Breeveld et al.(2005)]{Breeveld} Breeveld, A.~A., et 
al.\ 2005, \procspie, 5898, 391 

 \bibitem[Brown et al.(2005)]{brown05am} Brown, P. J. et al. \ 2005, \aj, 635, 1192
\bibitem[Brown et al.(2007)]{Brown05cs} Brown, P.~J., et al.\ 2007, \apj, 659, 1488 
\bibitem[Brown et al.(2008)]{BrownUV} Brown, P.~J., et al.\ 2008, ArXiv e-prints, 803, arXiv:0803.1265 
 
 \bibitem[Burket \& Li(2005)]{burket05hk}Burket, J., \& Li, W. \ 2005, IAU Circ., 8625
\bibitem[Burrows(1994)]{BurrowsHST} Burrows, C.~J.\ 1994, Hubble 
Space Telescope Handbook, V2
 \bibitem[Burrows et al.(2005)]{burrows05} Burrows, D. N., et al. \  2005, \ssr,
  120, 165
\bibitem[Campana et al.(2006)]{Campana} Campana, S., et al.\ 2006, \nat, 442, 1008 


\bibitem[Cappellaro et al.(1995)]{ulda} Cappellaro, E., Turatto, M., \& Fernley, J.\ 1995, ESA Special Publication, 1189

\bibitem[Chornock et al.(2006)]{Chornock05hk} Chornock, R., Filippenko, A.~V., Branch, D., Foley, R.~J., Jha, S., 
\& Li, W.\ 2006, \pasp, 118, 722 

  \bibitem[Crotts et al.(2006)]{crotts06jc} Crotts, A., Eastman, J., Depoy, D., Prieto, J.~L., \& Garnavich, P.\ 2006, CBET, 672, 1 
  \bibitem[Dessart \& Hillier(2005)]{Dessart2005} Dessart, L., \& Hillier, D.~J.\ 2005,  \aap, 437, 667
  \bibitem[Dessart \& Hillier(2006)]{Dessart2006} Dessart, L., \& Hillier, D.~J.\ 2006,  \aap, 447, 691 
 \bibitem[Dessart et al.(2008)]{DessartUV}  Dessart, L., et al.\ 2008, \apj, 675, 644 

\bibitem[Elmhamdi et al.(2003)]{E99em} Elmhamdi, A., et al.\ 
2003, \mnras, 338, 939 

 \bibitem[Ensman \& Burrows(1992)]{Ensman} Ensman, L., \& Burrows, A.\ 1992, \apj, 393, 742 

 \bibitem[Eastman \& Kirshner(1989)]{Eastman} Eastman, R.~G., \& Kirshner, R.~P.\ 1989, \apj, 347, 771 
\bibitem[Elias-Rosa,(2008)]{Elias_pro} Elias-Rosa, N,  2008, to appear in ``Supernovae: light in the darkness (XXIII Trobades cientifiques de la Mediterrania)'', 2007 October, Ma\'o, Menorca, Proceedings of Science
\bibitem[Ellis et al.(2008)]{Ellis08} Ellis, R.~S., et al.\ 
2008, \apj, 674, 51 
 


\bibitem[Evans (2005)]{Evans05df} Evans, R. , \ 2005, IAU Circ., 8584, 3
\bibitem[Fesen et al.(1999)]{Fesen79c} Fesen, R.~A., et al.\ 1999, \aj, 117, 725 
\bibitem[Fesen et al.(2006)]{fesen06jc} Fesen, R., Milisavljevic, D., \& Rudie, G.\ 2006, CBET, 672, 2 
\bibitem[Filippenko \& Chornock(2002)]{Filippenko02ao} Filippenko, A.~V., \& Chornock, R.\ 2002, \iaucirc, 7825, 1 

\bibitem[Foley et al.(2007)]{foley06jc} Foley, R.~J., Smith, N., Ganeshalingam, M., Li, W., Chornock, R., \& Filippenko, A.~V.\ 2007, \apjl, 657, L105 
\bibitem[Foley et al.(2008a)]{FoleyHZ} Foley, R.~J., et al.\ 2008, \apj, 684, 68 

\bibitem[Foley et al.(2008b)]{FoleyUV} Foley, R.~J., Filippenko, 
A.~V., \& Jha, S.~W.\ 2008b, \apj, 686, 117 

\bibitem[Fransson(1994)]{Fransson1994} Fransson, C.\ 1994, Supernovae, 677 

\bibitem[Fransson et al.(1984)]{Fransson1984} Fransson, C., Benvenuti, P., Wamsteker, W., Gordon, C., Hempe, K., Reimers, D., Palumbo, G.~G.~C., \& Panagia, N.\ 1984, \aap, 132, 1 
\bibitem[Fransson et al.(2002)]{Fransson2002} Fransson, C., et al.\ 2002, \apj, 572, 350 
\bibitem[Fransson et al.(2005)]{Fransson93J98S} Fransson, C., et al.\ 2005, \apj, 622, 991 
\bibitem[Gal-Yam et al.(2008)]{Gal-Yam} Gal-Yam, A., et al.\ 2008, \apjl, 685, L117  
\bibitem[Garavini et al.(2007)]{Garavini05cf} Garavini, G., et al.\ 2007,  \aap, 471, 527 
\bibitem[Gehrels et al.(2004)]{Gehrels} Gehrels, N., et al.\ 2004, \apj, 611, 1005 (erratum, 621, 558 [2005]) 
\bibitem[Gezari et al.(2008)]{Gezari08} Gezari, S., et al.\ 2008, \apjl, 683, L131 
 
\bibitem[Ghisellini et al.(2007)]{Ghisellini06aj} Ghisellini, G., 
Ghirlanda, G., \& Tavecchio, F.\ 2007, \mnras, 382, L77 
\bibitem[Hillebrandt \& Niemeyer(2000)]{hille00} Hillebrandt, W., \& Niemeyer, J.~C.\ 2000, \araa, 38, 191
\bibitem[Hoeflich et al.(1998)]{Hoeflich} Hoeflich, P., Wheeler, 
J.~C., \& Thielemann, F.~K.\ 1998, \apj, 495, 617 


\bibitem[Immler et al.(2005)]{Immler79C} Immler, S., et al.\ 2005, \apj, 632, 283 
 \bibitem[Immler et al.(2006a)]{Immler05ke} Immler, S., et al.\ 2006a, \apjl, 648, L119 
\bibitem[Immler(2006b)]{Immler06x} Immler, S.\ 2006b, The Astronomer's Telegram, 726, 1 
\bibitem[Immler et al.(2006c)]{ImmlerAtel06bp}Immler, S., Brown, P., \& Milne, P. \ 2006c, ATel, 793, 1 
\bibitem[Immler et al.(2007)]{Immler06bp} Immler, S., et al.\ 2007, \apj, 664, 435 
\bibitem[Immler et al.(2008)]{Immler06jc} Immler, S., et al.\ 2008, \apjl, 674, L85 
\bibitem[Jeffery et al.(1994)]{Jeffery1994} Jeffery, D.~J., et al.\ 1994, \apjl, 421, L27 

\bibitem[Kirshner et al.(1993)]{Kirshner92A} Kirshner, R.~P., et al.\ 1993, \apj, 415, 589 
\bibitem[Kloehr et al.(2005)]{Kloher} Kloehr W., Muendlein R., Li W., Yamaoka H., \& Itagaki K., \ 2005, IAU Circ., 8553, 1 
 
\bibitem[Leibundgut et al.(1991)]{Leibundgut90n} Leibundgut, B., Kirshner, R.~P., Filippenko, A.~V., Shields, J.~C., Foltz, C.~B., Phillips, M.~M., \& Sonneborn, G.\ 1991, \apjl, 371, L23 
\bibitem[Lentz et al.(2000)]{Lentz2000} Lentz, E.~J., Baron, E., Branch, D., Hauschildt, P.~H., \& Nugent, P.~E.\ 2000, \apj, 530, 966 
\bibitem[Lentz et al.(2001)]{Lentz2001} Lentz, E.~J., et al.\ 2001, \apj, 547, 406 
\bibitem[Levesque et al.(2005)]{Levesque} Levesque, E.~M., Massey, P., Olsen, K.~A.~G., Plez, B., Josselin, E., Maeder, A., \& Meynet, G.\ 2005, \apj, 628, 973 
\bibitem[Li et al.(2003)]{Li02cx} Li, W., et al.\ 2003, \pasp, 115, 453 


\bibitem[Lucy(1999)]{Lucy1999} Lucy, L.~B.\ 1999,  \aap, 345, 211 
\bibitem[Matheson et al.(2000)]{Matheson99cq} Matheson, T., Filippenko, A.~V., Chornock, R., Leonard, D.~C., \& Li, W.\ 2000, \aj, 119, 2303 
\bibitem[Martin et al.(2005)]{Martin05am} Martin, R., \ 2005, IAU Circ. 8490, 1
 \bibitem[Mason et al.(2001)]{MasonXXM} Mason, K.~O., et al.\ 2001,  \aap, 365, L36 

\bibitem[Mazzali et al.(1993)]{Mazzali90N} Mazzali, P.~A., Lucy, L.~B., Danziger, I.~J., Gouiffes, C., Cappellaro, E., \& Turatto, M.\ 1993, \aap, 269, 423 
\bibitem[Mazzali \& Lucy(1993)]{MazzaliLucy93} Mazzali, P.~A., \& Lucy, L.~B.\ 1993,  \aap, 279, 447 
\bibitem[Mazzali \& Chugai(1995)]{MazzaliChugai} Mazzali, P.~A., \& Chugai, N.~N.\ 1995,  \aap, 303, 118 

\bibitem[Mazzali(2000)]{Mazzali2000} Mazzali, P.~A.\ 2000,  \aap, 363, 705 
\bibitem[Mazzali et al.(2005)]{MazzaliHVF} Mazzali, P.~A., et al.\ 
2005, \apjl, 623, L37 
\bibitem[Mazzali et al.(2008)]{Mazzali08D} Mazzali, P.~A., et al.\ 
2008, Science, 321, 1185   

\bibitem[Milliard et al.(2001)]{MilliardGALEX} Milliard, B., et al.\ 2001, Mining the Sky, 201 

 \bibitem[Modjaz et al.(2005a)]{Modjaz05am} Modjaz, M., Kirshner, R., \& Challis, P. \ 2005a, IAU Circ. 8491, 2 
\bibitem[Modjaz et al.(2005b)]{Modjaz05cf} Modjaz, M., Kirshner, R., Challis, P., \& Berlind, P. \ 2005b,  IAU Circ., 8534, 3 
\bibitem[Modjaz et al.(2005c)]{Modjaz05cs}Modjaz M., Kirshner R., Challis P.,  \& Hutchins R. \ 2005c, IAU Circ., 8555, 1
\bibitem[Modjaz et al.(2006)]{Modjaz06jc} Modjaz, M., Blondin, S., Kirshner, R., Challis, P., Matheson, T., \& Mamajek, E.\ 2006, CBET, 677, 1 
\bibitem[Modjaz et al.(2008)]{Modjaz08D} Modjaz, M., et al.\ 2008, arXiv:0805.2201 


\bibitem[Nakano et al.(2006a)]{nakano06bp} Nakano, S. \ 2006a, IAU Circ., 8700, 2
\bibitem[Nakano et al.(2006b)]{nakano06jc}  Nakano, S., Itagaki, K., Puckett, T., \& Gorelli, R. \ 2006, CBET, 666 
\bibitem[Panagia et al.(1980)]{Panagia1980} Panagia, N., et al.\ 
1980, \mnras, 192, 861 
\bibitem[Panagia(1985)]{Panagia1985} Panagia, N.\ 1985, Supernovae as Distance Indicators, Lecture
Notes in Physics, 224, 14 
\bibitem[Panagia(2003)]{Panagia2003} Panagia, N., \ 2003,  in "Supernovae and Gamma-Ray Bursters", ed. K.W.Weiler (Springer-Verlag:Berlin), p.113-114 
\bibitem[Panagia(2007)]{Panagia2007} Panagia, N.\ 2007, American Institute of Physics Conference Series, 937, 236

\bibitem[Pastorello et al.(2006)]{pastorello05cs}  Pastorello, A., et al. \ 2006, \mnras, 370, 1752 
\bibitem[Pastorello et al.(2007a)]{pastorello05cf} Pastorello, A. \ 2007a, \mnras, 376, 1301
\bibitem[Pastorello et al.(2007b)]{pastorello06jc} Pastorello, A., et al.\ 2007b, \nat, 447, 829 
\bibitem[Pastorello et al.(2008)]{Pastorellofamily} Pastorello, A., et al.\ 2008, \mnras, 389, 113 

\bibitem[Patat et al. (2005)]{patat05ke} Patat, F., Baade, D., Wang, L., Taubenberger, S., \& Wheeler, J. C. \ 2005, IAU Circ. 8631 
\bibitem[Patat et al.(2007)]{Patat06x} Patat, F., et al.\ 2007, 
Science, 317, 924 
\bibitem[Pauldrach et al.(1996)]{Pauldrach96} Pauldrach, A.~W.~A., Duschinger, M., Mazzali, P.~A., Puls, J., Lennon, M., \& Miller, D.~L.\ 1996,  \aap, 312, 525 

\bibitem[Phillips et al.(2007)]{phillips05hk} Phillips, M.~M., et al.\ 2007, \pasp, 119, 360

\bibitem[Poole et al.(2008)]{PooleUVOT} Poole, T.~S., et al.\ 2008, \mnras, 383, 627 

\bibitem[Pugh \& Li(2005)]{Puckett05cf} Pugh, H., \&  Li, W., \ 2005, IAU Circ., 8534, 1

\bibitem[Pun et al.(1995)]{Pun87A} Pun, C.~S.~J., et al.\ 1995, \apjs, 99, 223 

\bibitem[Quimby et al.(2006a)]{Quimby06x} Quimby, R., Brown, P., Gerardy, C., Odewahn, S.~C., 
\& Rostopchin, S.\ 2006, CBET, 393, 1 
\bibitem[Quimby et al.(2006b)]{QuimbyCBET06bp} Quimby, R., Brown, P., Caldwell, J., \& Rostopchin, S.\ 2006b, Central Bureau Electronic Telegrams, 471, 1 

\bibitem[Quimby et al.(2007)]{Quimby06bp} Quimby, R.~M., Wheeler, J.~C., H{\"o}flich, P., Akerlof, C.~W., Brown, P.~J., \& Rykoff, E.~S.\ 2007, \apj, 666, 1093 

 
\bibitem[Roming et al.(2005)]{roming05} Roming, P. W. A. et al. \ 2005, \ssr,
120,95

\bibitem [Salvo \& Schmidt (2005)]{salvo}Salvo, M., \& Schmidt, B. \ 2005, IAU Circ. 8581, 2
\bibitem[Sauer et al.(2008)]{sauer08}Sauer, D.~N., et al.\ 2008, \mnras, 391, 1605   

\bibitem[Sahu et al.(2007)]{Sahu05hk}  Sahu, D.~K., et al.\ 2008, 
\apj, 680, 580 
\bibitem[Schawinski et al.(2008)]{Schawinski} Schawinski, K., et al.\ 2008, Science, 321, 223 

\bibitem[Schlegel et al.(1998)]{Schlegel} Schlegel, D.~J., Finkbeiner, D.~P., \& Davis, M.\ 1998, \apj, 500, 525 

\bibitem[Serduke et al.(2005)]{serduke} Serduke, F.~J.~D., Wong, D.~S., \& Filippenko, A.~V.\ 2005, CBET, 269, 1 
\bibitem[Soderberg et al.(2008)]{Soderberg}  Soderberg, A.~M., et al.\ 2008, \nat, 453, 469 

\bibitem[Stanishev et al.(2007)]{Vallery05hk} Stanishev, V., et 
al.\ 2007, American Institute of Physics Conference Series, 924, 336 

\bibitem[Suzuki \& Migliardi(2006)]{circ06x}Suzuki, S., \& Migliardi, M. \ 2006, \iaucirc, 8667, 1 

\bibitem[Taubenberger et al.(2008)]{Tau05bl} Taubenberger, S., et al.\ 2008, \mnras, 385, 75 
\bibitem[Turatto(2003)]{Turattoclass} Turatto, M.\ 2003, Supernovae 
and Gamma-Ray Bursters, 598, 21 

\bibitem[Yamaoka et al.(2005)]{Yamaoka05am} Yamaoka, H., \ 2005. IAU Circ. 8490, 2
\bibitem[Wang et al.(2008a)]{Wang06x} Wang, X., et al.\ 2008a, \apj, 675, 626 
\bibitem[Wang et al.(2008b)]{Wang05cf} Wang, X., et al.\ 2008b, arXiv:0811.1205 


 
\bibitem[Wheeler et al.(1986)] {wheeler}Wheeler, J. C., et al. \ 1986, PASP, 98, 1018
\bibitem[Waxman et al.(2007)]{Waxman} Waxman, E., M{\'e}sz{\'a}ros, P., \& Campana, S.\ 2007, \apj, 667, 351 

\end{thebibliography}
\end{document}